\documentclass[preprint2]{emulateapj}
\usepackage{epsfig}
\usepackage{amssymb}
\usepackage{graphicx}
\usepackage{color}
\usepackage{epstopdf}
\usepackage{subfigure}
\usepackage{natbib}

\slugcomment{In preparation}
\shorttitle{Coronal heating through footpoint braiding}
\shortauthors{Guerreiro et al.}
\begin{document}

\title{Numerical simulations of coronal heating through footpoint braiding}

\author{ V. Hansteen}
\affil{Institute of Theoretical Astrophysics, University of
  Oslo, P.O. Box 1029 Blindern, N-0315 Oslo, Norway}
\email{viggo.hansteen@astro.uio.no}

\author{N. Guerreiro}
\affil{Physikalisch-Meteorologische Observatorium, PMOD/WRC, Dorfstrasse 33, 7260 Davos Dorf, Switzerland}

\author{ B. De Pontieu}
\affil{Lockheed Martin Solar and Astrophysics Laboratory, 3251 Hanover Street, Organization A021S, Building 252, Palo Alto, CA 94304, USA}
\affil{Institute of Theoretical Astrophysics, University of
  Oslo, P.O. Box 1029 Blindern, N-0315 Oslo, Norway}

\author{M. Carlsson}
\affil{Institute of Theoretical Astrophysics, University of
  Oslo, P.O. Box 1029 Blindern, N-0315 Oslo, Norway}

\begin{abstract}
Advanced 3D radiative MHD simulations now reproduce many properties of the outer solar atmosphere. When including a domain from the convection zone into the corona, a hot chromosphere and corona are self-consistently maintained. Here we study two realistic models, with different simulated area, magnetic field strength and topology, and numerical resolution. These are compared in order to characterize the heating in the 3D-MHD simulations which self-consistently maintains the structure of the atmosphere. We analyze the heating at both large and small scales and find that heating is episodic and highly structured in space, but occurs along loop shaped structures, and moves along with the magnetic field. On large scales we find that the heating per particle is maximal near the transition region and that widely distributed opposite-polarity field in the photosphere leads to a greater heating scale height in the corona. On smaller scales, heating is concentrated in current sheets, the thicknesses of which are set by the numerical resolution. Some current sheets fragment in time, this process occurring more readily in the higher-resolution model leading to spatially highly intermittent heating. The large scale heating structures are found to fade in less than about five minutes, while the smaller, local, heating shows time scales of the order of 2 minutes in one model and 1 minutes in the other, higher-resolution, model. 
\end{abstract}


\section{INTRODUCTION}

The last few decades have seen great progress, both observationally and theoretically, in understanding the heating of the solar chromosphere and corona. These advances have occured through the Solar and Heliospheric Observatory \citep{V.Domingo121995}, Transition Region and Coronal Explorer \citep{B.Handy071999}, Hinode \citep{T.Kosugi062007}, Solar Dynamics Observatory/Atmospheric Imaging Assembly \citep{2012SoPh..275...17L}, Interface Region Imaging Spectrograph \citep{2014SoPh..289.2733D} satellites, and ground-based observatories and instruments such as the Swedish 1-m Solar Telescope \citep{2003SPIE.4853..341S,2006AA...447.1111S,2008ApJ...689L..69S}, the Interferometric Bidimensional Spectrometer ( \citeauthor{2006SoPh..236..415C}, \citeyear{2006SoPh..236..415C}) at the Dunn Solar Telescope and GREGOR \citep{2012AN....333..796S}. This wealth of high quality data has been matched by the development of sophisticated  numerical models of the outer solar atmosphere \citep[e.g.][]{B.Gudiksen012005,S.Bingert062011,2007ApJ...665.1469A}. The convection zone clearly contains sufficient mechanical energy to heat the corona. However, the mechanisms responsible for the transport and release of the energy from the photosphere through the chromosphere to the corona remain elusive \citep[see, e.g.][]{R.Walsh122003,J.Klimchuk032006,2012RSPTA.370.3217P}.

Observational constraints \citep{R.Athay121978,R.Athay051979,N.Mein041981,A.Fossum062005,M.Carlsson112007}
as well as numerical models \citep{M.Carlsson062002} indicate that the power contained in acoustic waves is not sufficient to heat or maintain a chromosphere or corona as found on the Sun except perhaps in regions of very weak magnetic fields. Therefore, 
most coronal heating models require that the magnetic field plays an important role in the processes of transport, storage, and release of energy.

Several models have been proposed to solve these problems. These can be subdivided into two classes involving either wave heating, ``AC''  models, where the magnetic field is stressed at relatively high frequency {compared to chromospheric or coronal dynamical time scales},  or nano-flare ``DC'' models, where magnetic field stresses are built up over longer time scales \citep{J.Klimchuk032006}. 
Both types of models face constraints set by the nature of the solar atmosphere. The AC models must explain the difficulties of transporting wave energy through the chromosphere, as fast mode waves will be refracted towards regions of low Alfv\'en speed. Alternately, if Alfv\'en waves are considered, the expected low dissipation rates of such waves in the corona must be overcome \citep{A.Bondeson001985,A.Ballegooijen072011,M.Asgari022012}. DC models must explain the formation of sufficiently large magnetic field gradients in the corona to give sufficient heating in the face of the intrinsically low coronal resistivity. For a comprehensive discussion about the magnetic models we refer to e.g. \citet{J.Klimchuk032006}.

In this paper we concentrate on the DC heating type models and consider the properties of numerical simulations of nano-flare heating. { Though several such models have been presented previously \citep{B.Gudiksen012005,M.Berger112009,S.Bingert062011},
these lacked the self-consistent convective motions included in our
simulations.} We believe that the models presented here are the first
that demonstrate that the chromospheric and coronal magnetic field
gradients built up through such self-consistent convection are
sufficient to maintain coronal temperatures over an extended period. In particular, we study how the turbulent motions in the convection zone and photosphere drag and twist the magnetic field. This leads to a Poynting flux into the corona, with steadily increasing free magnetic energy stored in the field. 
This continuing turbulent motion increases the stored energy in the field until the maximum storage capability is reached and the magnetic gradients required for free energy storage no longer can be supported. A quasi steady-state ensues where the energy injected is balanced by dissipation at small scales.

Thus, dissipation occurs as the magnetic field attains sufficient gradients to generate small scale current sheets which, by reconnection, release the stored magnetic energy and create a simpler magnetic topology as described by \citet{K.Galsgaard061996}. In general, energy is therefore released in the form of resistive currents, plasma flows, { wave excitation}, particle acceleration, or through viscosity as plasma flows and other motions are thermalized \citep{E.Priest2000,G.Baumann002012a}. 

In the MHD type of model described here we do not have sufficient spatial resolution nor include enough microphysics to model all these aspects of magnetic reconnection in the corona with high fidelity. The basic assumption is therefore that on the scale we are modeling the system, the microscopic description of these processes is not vital and that the large scale evolution of the system is insensitive to the details of how magnetic energy is thermalized. There is mounting evidence that this assumption is not unreasonable \citep{K.Galsgaard061996,B.Gudiksen012005,M.Berger112009,S.Bingert062011,D.Pontin052011}. 
 
Though they do not have the spatial resolution to rule out that AC heating plays a significant role in coronal heating, the recent development of ``realistic'' 3D MHD models spanning from the photosphere to the corona \citep{B.Gudiksen012005} support the nanoflare  theory. These models show intense Joule heating events or the formation of current sheets throughout the chromosphere all the way to the corona where the Joule heating per particle peaks around the upper chromosphere and lower transition region \citep{V.Hansteen082010}. 
In summary, large scale photospheric motions drive the magnetic field gradients in the chromosphere and corona to small scales, thereby forcing episodic dissipation at the same small scales. These dissipation events (nanoflares) were first put forward by 
\cite{R.Levine061974}
as a mechanism of heating the corona. The nanoflare theory was further elaborated by \cite{E.Parker011983,E.Parker071988,P.Cargill021994,J.Klimchuk052001,E.Priest092002}.

Parker estimated an energy of about $10^{16}$ to $10^{17}$ J per event, and noted that a sufficient number of events, when combined over the Sun, should be able to provide most of the energy needed to explain the coronal energy budget \citep{E.Parker071988}. 

Despite the success of numerical models, observational confirmation is of course critical. Several studies \citep{D.Datlowe111974,R.Lin081984,B.Dennis1985,N.Crosby021993} suggest that solar flares occur with an energy distribution given by the power law $dN/dE\sim E^{-\alpha}$ with an index $\alpha \sim 1.8$, where $N$ is the number of events and $E$ is the energy of each event. However, if the nanoflares are to be energetically important to coronal heating $\alpha$ is required to be $>2$ \citep{H.Hudsen061991}. 

Several statistical studies \citep{S.Krucker071998,D.Berghmans081998,C.Parnell012000,M.Aschwanden062000,A.Benz2002} have attempted to compute the value of $\alpha$ for EUV, small-scale transient brightening, or nanoflare-like heating events, across the solar disk, but have so far proved inconclusive. The models used in those studies relay upon iron lines (at 171 \AA~and 195~\AA) to detect and diagnose the events. \cite{S.Krucker071998} developed a method to assess heating events by calculating the emission measure at all pixels for the entire time series. This model gives a power-law with an index $\alpha $ that is greater than 2. \cite{M.Aschwanden062000} on the other hand, present a more restrictive method in selecting heating events; they use a pattern recognition code that extracts spatiotemporal events with significant variability. This method gives a power law with an index $\alpha $ that is less than 2.  The difficulty in reaching an agreement on the frequency and energetics of nanoflares arises from the difficulty in assessing when an event has happened. Without such agreement it is difficult to estimate whether the contribution from nanoflares is of the same order as the $300~\rm W~m^{-2}$ that is required to heat the quiet corona \citep{G.Withbroe001977}. 

There are also other observational puzzles that may be shedding light on the nature of coronal heating. Amongst these are the nature of the spicule acceleration and heating mechanism \citep{B.Pontieu082009,B.Pontieu012011}, the run of the differential emission measure with temperature \citep{2006ApJ...638.1086P}, and the ubiquitous average redshift seen in lower transition region emission lines \citep{H.Peter091999}. A successful model for coronal heating should be able to explain how these phenomena arise and predict how they vary at differing magnetic topologies and field strengths \citep[e.g.][]{V.Hansteen082010,2015ApJ...802....5O}.

In this paper we expand on the work done by \citet{V.Hansteen082010} and characterize Joule heating events in two models with different initial magnetic field distributions and different spatial resolutions. We will discuss the shapes, lifetimes and sizes of the current sheets together with the structure of the magnetic field lines about the current sheets. 

A short description of the code used and the models are presented in section $\S $ 2, the characterization of the heating events is given in $\S $ 3, and the discussion and conclusion follows in section $\S $ 4.

\section{NUMERICAL METHOD AND MODEL}

We wish to model coronal heating resulting from convective motions by consistently simulating the layers of the solar atmosphere from the upper convection zone to the corona. Solar convection is maintained by setting the entropy of inflowing material at the bottom boundary of the computational domain and solving the equations of radiative transfer in four frequency bins such that an atmosphere with effective temperature approximately equal to the solar $T_{\rm eff}=5780$~K arises. This is a well known technique \citep{A.Nordlund031982,M.Steffen031988}
that reproduces solar photospheric intensities, energetics, and dynamics to a high degree of precision, seemingly only limited by the spatial resolution of the model. 

In order to extend the model into the chromosphere we include the effects of scattering in the equations of radiative transfer following \cite{R.Skartlien062000}.

A magnetic field must also be introduced into the model. This is done by setting the vertical magnetic field at the bottom boundary at time $t=0$~s and extending the field into layers above by performing a potential field extrapolation. As the model evolves we allow the field to move with the fluid motions at the bottom boundary. In the models presented here no new flux is injected into the bottom boundary during the simulation run. 

The model is evolved in time by solving the radiative 3D MHD equations using the Bifrost code. A description of the code and treatment of the MHD equations can be found in \cite{B.Gudiksen052011}. In summary, the code includes artificial viscous diffusivity and magnetic resistivity terms, which are the sources of viscous and magnetic heating.  Non-grey, non-LTE radiative transfer in the photosphere, and optically and effectively thin radiation in the upper chromosphere are included following the schemes described above \citep{W.Hayek072010,M.Carlsson032012}.  The equation of state is given by a set of look up tables that given the internal energy and density return the pressure, temperature, opacities and other radiation quantities as described in \cite{B.Gudiksen052011}. Thermal conduction along the magnetic field lines is included and is implemented using an implicit algorithm solved using a multi grid method. The boundaries are periodic in the $x$ and $y$ direction, while being non-periodic in the vertical $z$ direction. The lower boundary allows flows to exit unimpeded while the entropy of inflowing material is set to maintain the effective temperature at $T_{\rm eff} \approx 5800$~K as mentioned above. The upper boundary is transparent. 

In both the two models (model A and model B) presented in this paper we find a corona characterized by plasma at $T>10^6$~K maintained ultimately by Joule heating events that either heat the plasma directly, or induce small scale shear flows that are thermalized via viscous dissipation, as the initially potential magnetic field is stressed by convection zone and photospheric motions. 

The two models differ in the grid resolution and initial magnetic field distribution. The model A described below is essentially the same model as the `B1' model described in \citet{V.Hansteen082010} and \citet{N.Guerreiro052013}.

\subsection{MODEL A}

The model A has a computational box with physical dimensions $16.6\times 8.3\times 15.5$~Mm on a grid of $256\times128\times160$ points. The grid is uniformly spaced, about $65$~km, in the $x$ and $y$ direction and non uniform in the $z$ direction, with $\Delta z$ ranging from $32$~km in the photosphere to $44$~km in the corona. We have used the convention of setting the zero point in our height scale in the photosphere at the point where on average $\tau_{500~{\rm nm}}=1$. The unsigned magnetic field strength is fairly large, of order 135~Gauss, but well mixed, with a net magnetic flux close to zero. Thus, this model could be typical of a strong-field, quiet-sun region with no field organized on larger than photospheric scales.

\begin{figure*}
\includegraphics{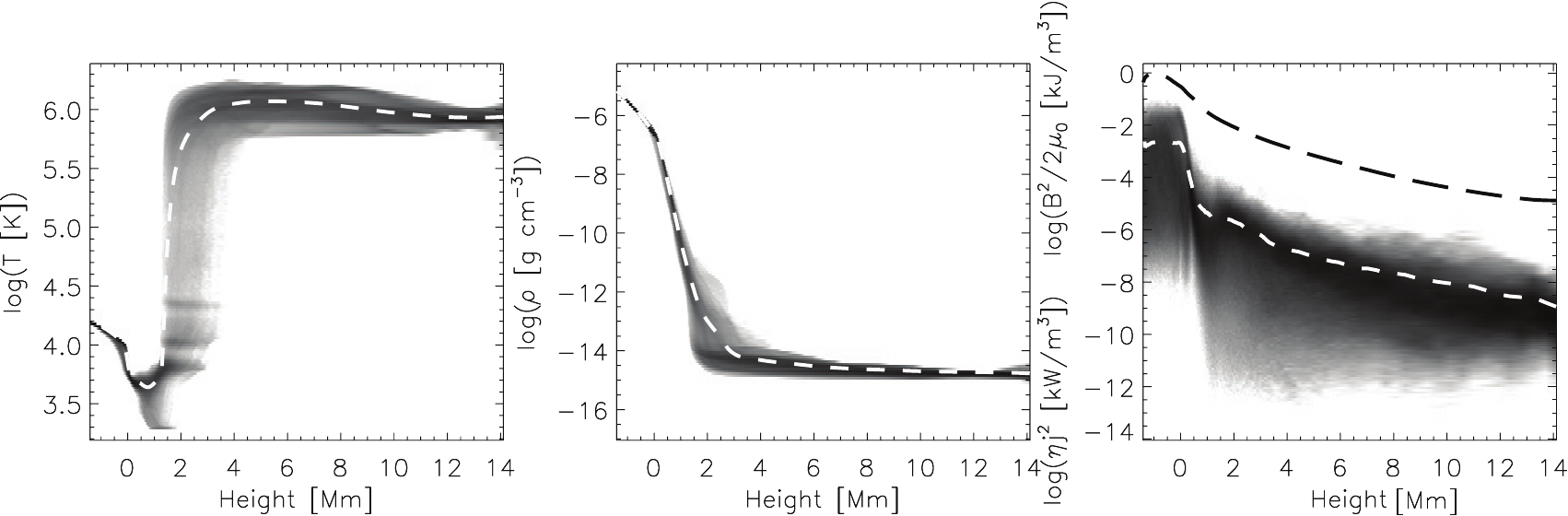}
 \caption{State of the atmosphere at $t=1000$~s as function of height in model A. From left to right: temperature, density and Joule heating. The white dashed lines are the variables average as a function of  height. The average magnetic energy is given in the rightmost panel as a dashed black line.} 
 \label{fig:histem_den_heatlr}
\end{figure*}

A summary of the state of the atmosphere at $t=1000$~s from the photosphere and up as function of height is shown in figure~\ref{fig:histem_den_heatlr}. The left panel shows the temperature distribution. In the photosphere,  the temperature varies between $5000$~K and $7500$~K, at $z=0~\rm Mm$. In the first 1000~km above the photosphere the average and minimum temperatures fall, while the maximum temperature falls to $5800$~K some $350$~km above the photosphere and increases thereafter.  At $1~\rm Mm$ we find that in regions of vigorous expansion, such as in the wake of shock waves, the minimum temperature falls to  $2000$~K. In order to avoid problems with the equation of state an artificial heating term is turned on at temperatures lower than $2000$~K \citep[see][]{J.Leenaarts062011}. The maximum temperature at 1~Mm reaches $6600$~K while the average is $4500$~K. From this height and up to the transition region the minimum temperature remains (artificially) fixed near $2000$~K, while the average and maximum temperature increase slowly. Material at transition region temperatures is found from $z=1.2$~Mm and up to some $3.5$~Mm. Above this height material is found in the temperature range $5\times 10^5$~K to $2\times 10^6$~K. Note the three ionization bands with elliptical shape corresponding to the ionization of H~{\sc i}, He~{\sc i}, and He~{\sc ii} from bottom to top, respectively \citep[see][]{J.Leenaarts062011}. We also find a large spread in the density in the upper chromosphere, transition region and lower corona between $1.5$ and $7$~Mm as shown in the middle panel of figure~\ref{fig:histem_den_heatlr}. At greater heights we find that the density is roughly constant; $\log_{10}\rho=-15.3\rm~g~cm^{-3}$ equivalent to electron densities slightly less than $n_{\rm e}=10^9\rm~cm^{-3}$ in the corona. 
The right panel of figure~\ref{fig:histem_den_heatlr} shows the Joule heating and the average magnetic energy density. We note that the ratio of gas pressure to magnetic pressure $\beta<1$ above some $1$~Mm. The Joule heating decreases exponentially with height from the middle chromosphere to the corona with a scale height that is closely related to the magnetic energy ($B^{2}/2\mu_{0}$) scale height, roughly  $\sim 1300$~km in this model. Similar findings were made by  \citet{B.Gudiksen012005} and \citet{V.Hansteen082010}. There is a large scatter in the strength of the Joule heating at any given height, up to 8 orders of magnitude.

This simulation starts with a semi-random distribution of vertical magnetic field ``poles'' at the bottom boundary designed to give a ``reasonable'' unsigned magnetic flux and a few large scale unipolar concentrations in the photosphere. The initial field in the computational box is computed through a potential field extrapolation. The simulation is run for over an hour solar time, during which the magnetic field is stressed by convective and photospheric motions.  The average field strength in the photosphere is of order $100$~Gauss, starting at $150$~Gauss at $t=0$~s and falling to $100$~Gauss at the end of the run as magnetic field is pulled down by convective motions. On the other hand, in the 
upper chromosphere and in the corona the mean unsigned field strength is very nearly constant or even increasing; for example at a height of 1.5~Mm above the photosphere we find an average field strength that varies in the range $[55,60]$~Gauss, while in the corona at a height of 4~Mm it is in the range $[15,18]$~Gauss. 

\begin{figure*}[!t]
\begin{center}
  \subfigure{\includegraphics[width=0.49\textwidth]{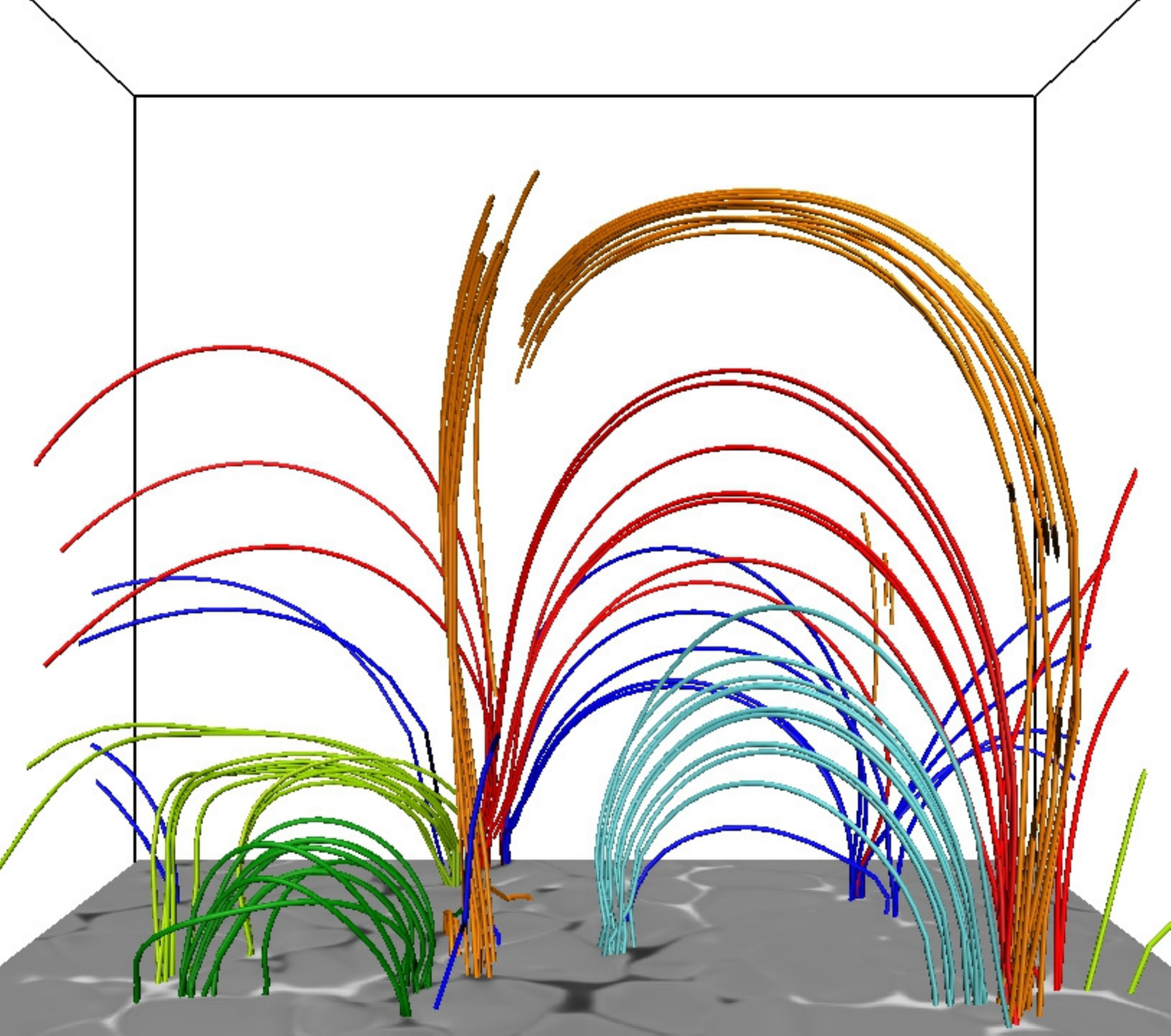}} 
 \subfigure{\includegraphics[width=0.49\textwidth]{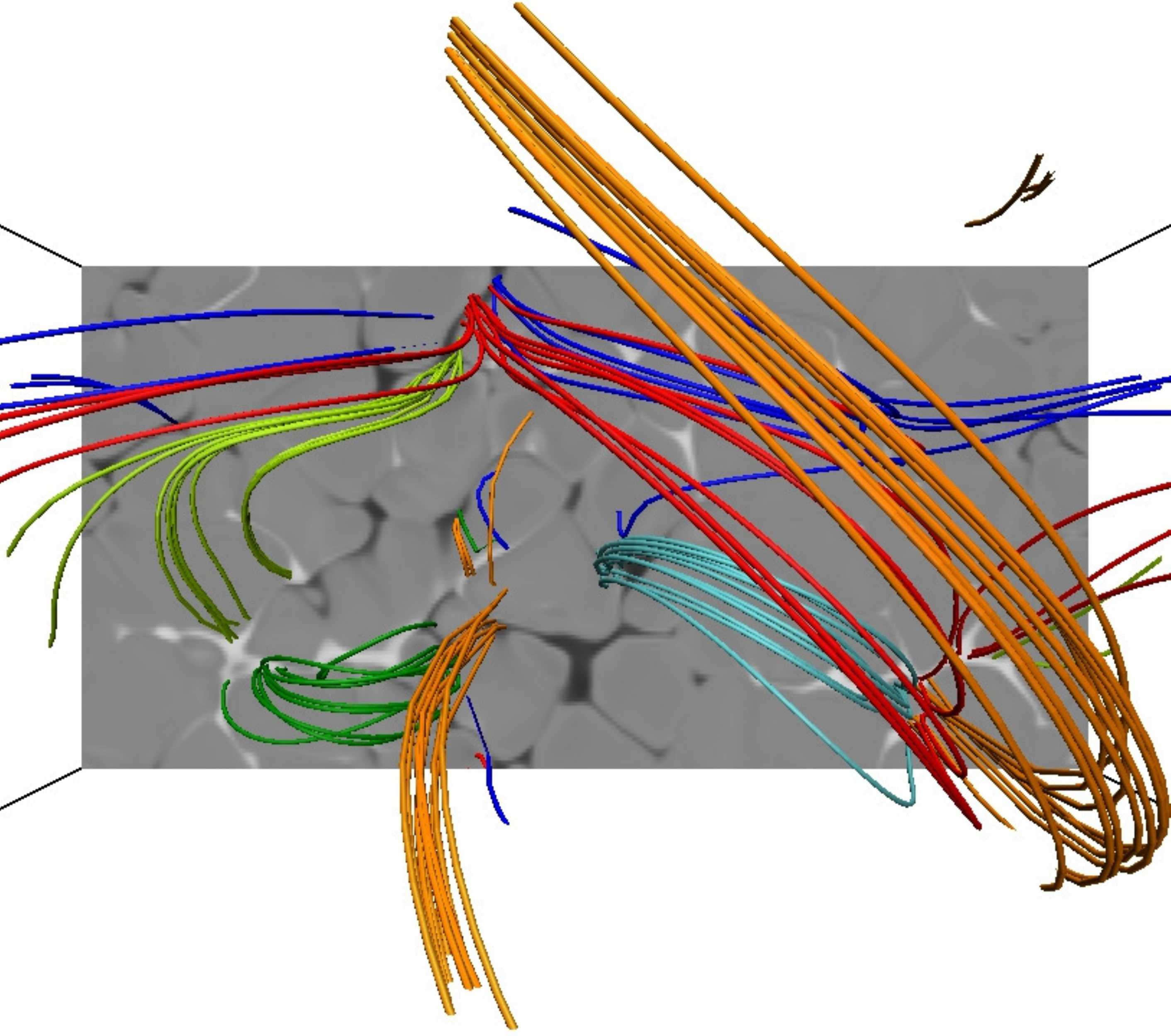}}
   \subfigure{\includegraphics[width=0.49\textwidth]{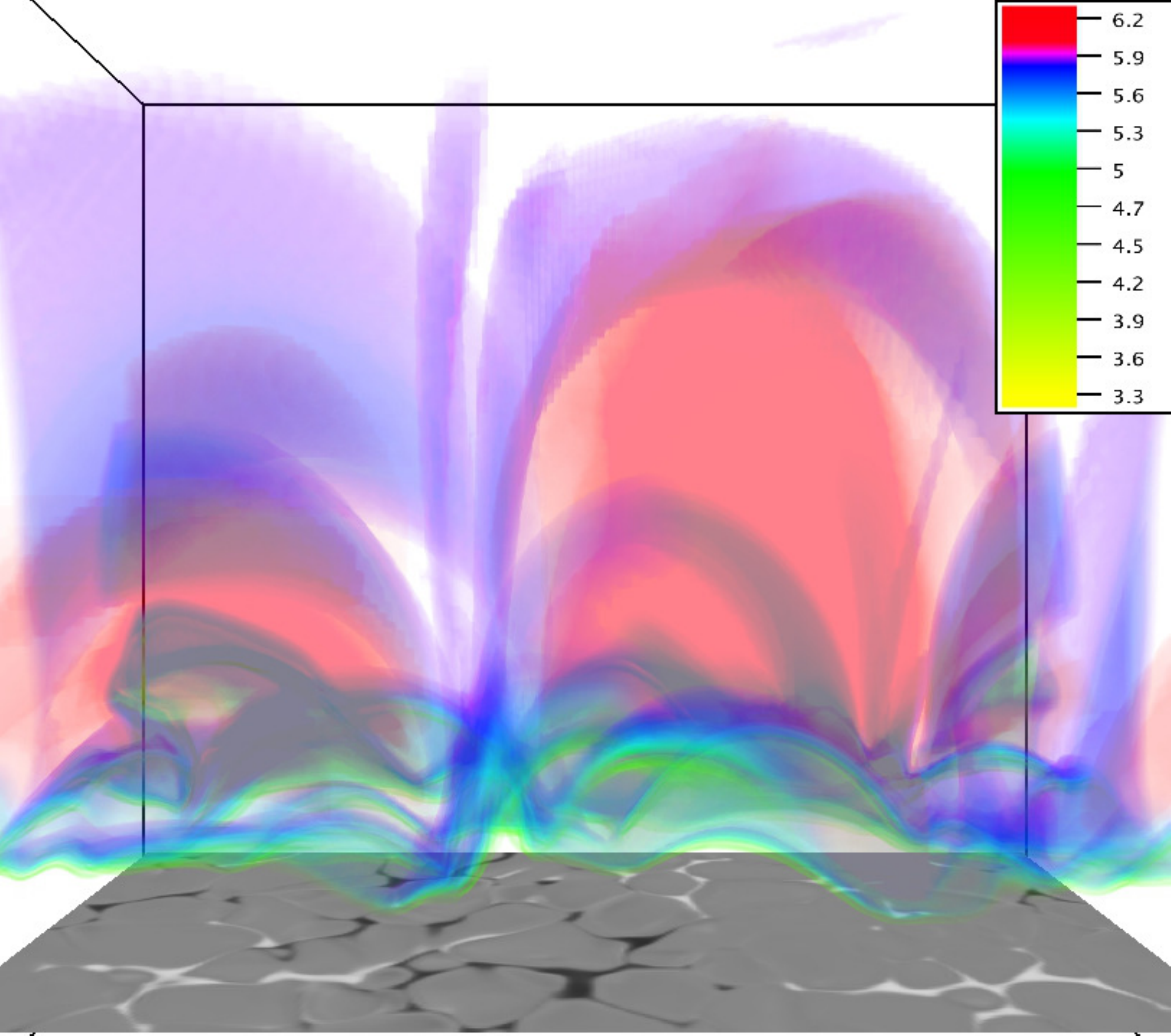}} 
  \subfigure{\includegraphics[width=0.49\textwidth]{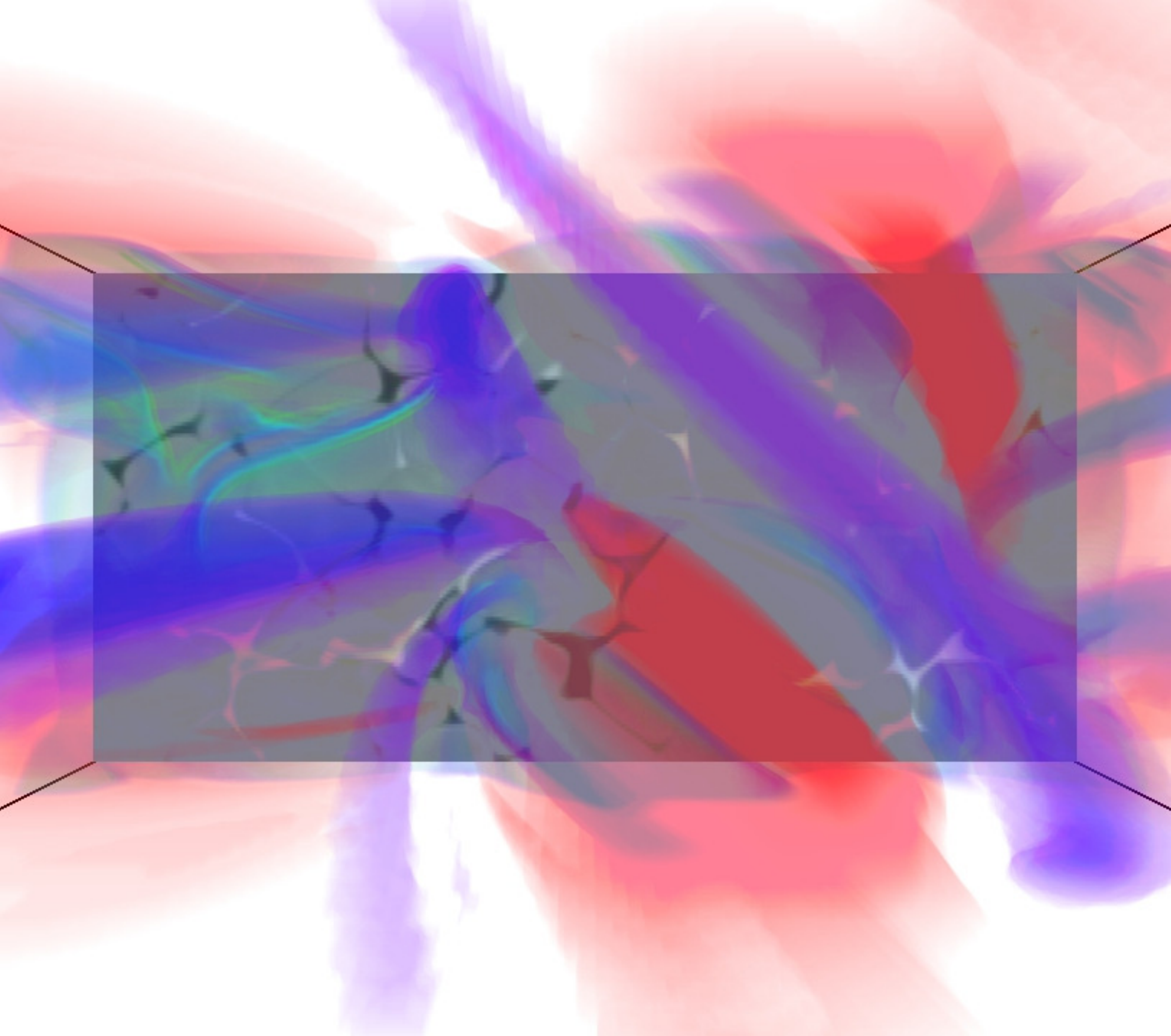}}
 \caption{Magnetic field configuration shown for loops with apex at different heights (top panels: side view-left, top view-right). Temperature stratification (bottom panels: side view-left, top view-right). The field lines have been given different colors according to their apex height above the photosphere. The temperature color scale is green and turquoise for the upper chromosphere and low transition region temperature, blue for the upper transition region temperature, purple for the low corona and red for the hottest coronal regions. The $B_{z}$ magnetic field component, (black negative polarity, white positive polarity) is shown in the photosphere at $z = 0~\rm Mm$ where $\tau_{500}=1$. These representations are for Model A.} 
 \label{fig:MF_modelslr}
\end{center}
\end{figure*}

In figure \ref{fig:MF_modelslr} we show the vertical field in the photosphere and a number of field lines extending into the corona at 8~minutes into the run.  The magnetic field configuration can also be recognized in the temperature structure of the corona where we see large loop shaped regions of nearly isothermal material following the magnetic field lines. The atmospheric structure is complex and there are large horizontal variations found, starting in the upper chromosphere and  transition region.

\begin{figure*}[!t]
\begin{center}
\subfigure{\includegraphics[width=0.49\textwidth]{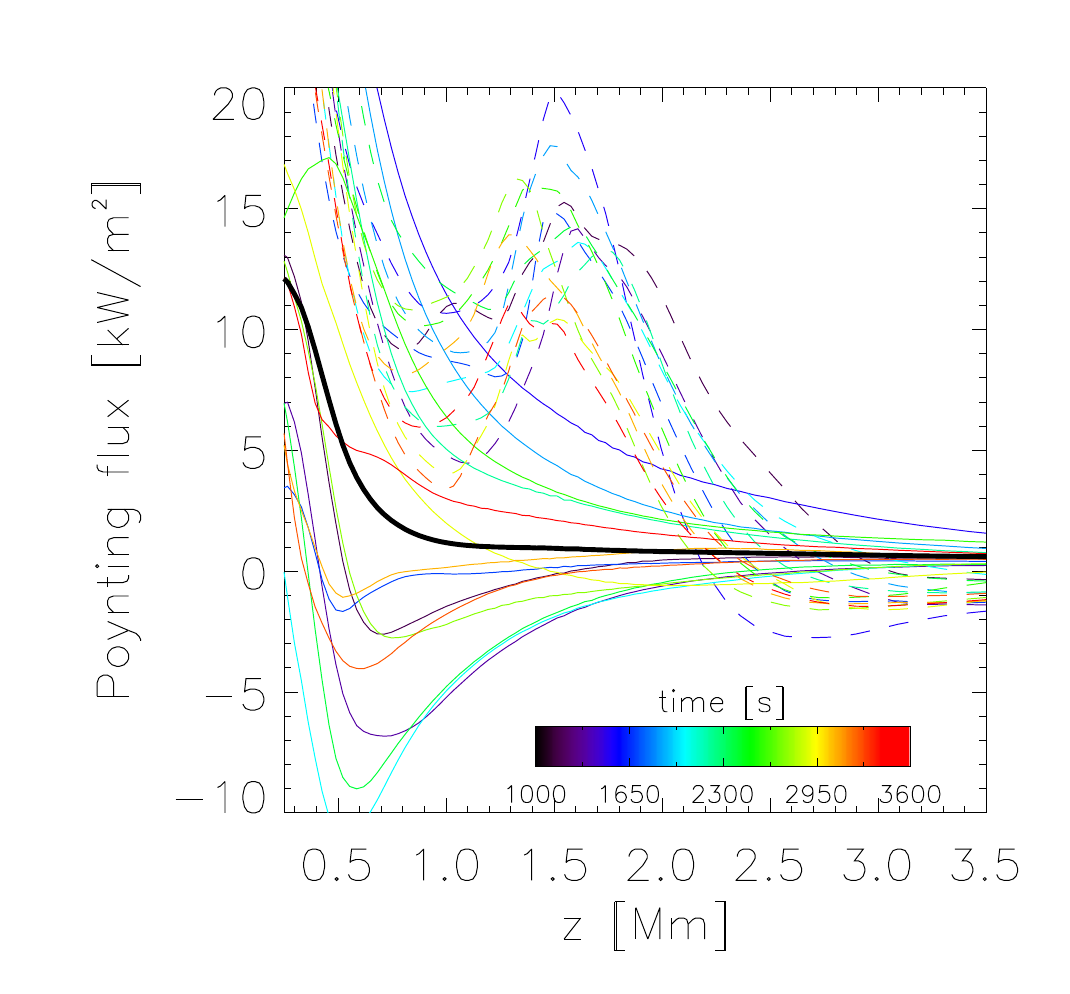}} 
\subfigure{\includegraphics[width=0.49\textwidth]{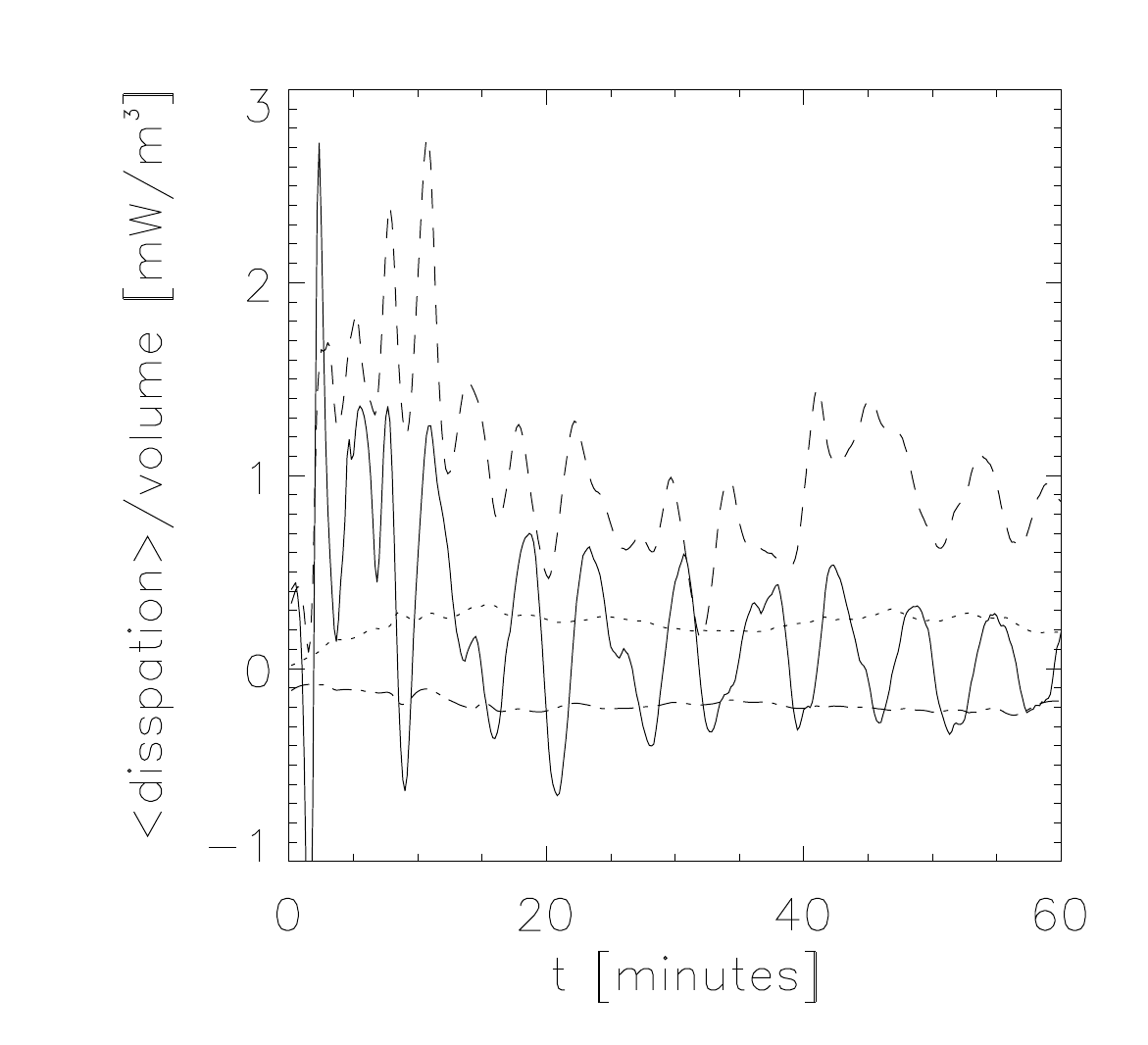}} 
 \caption{Left panel: The Poynting flux as a function of height every 200~s from $t=1000$~s to $t=3600$~s (colored lines from purple through blue, turquoise, yellow to red) and time average (thick black line) for the low resolution model. Right panel: Poynting flux at $z=1.3$~Mm divided by the distance of this height to the top of the simulation, along with the total spatially averaged Joule dissipation $\eta j^2$ (dotted line) and total spatially averaged radiative losses (dot-dashed line) above this height. In both panels the dashed lines show the vertical Poynting flux associated with horizontal motions, while the solid lines show the total vertical Poynting flux. These plots use data from model A.} 
 \label{fig:poynting_flux}
\end{center}
\end{figure*}

The energy flux that ultimately heats the upper chromosphere, transition region and corona is mediated from the photosphere in the form of a Poynting flux $\left({\mathbf E}\times{\mathbf B}\right)/\mu_{0}$. The vertical component of this flux can be divided into two parts; one is due the work done by horizontal motions on the vertical field
\begin{equation}\label{eq:poynting1}
-\frac{1}{\mu_{0}}B_z(u_yB_y+u_xB_x),
\end{equation}
the other due the transport of horizontal field by vertical motions
\begin{equation}\label{eq:poynting2}
\frac{1}{\mu_{0}}u_z(B_x^2+B_y^2),
\end{equation}
\noindent
but note that this division, perhaps convenient for observers, also contains contributions from the effects of waves and large terms of opposite sign, that will ultimately cancel, due to motions parallel to a tilted field.
Energy is transported both as a result of the horizontal motions associated with convection and granulation and as horizontal field is advected in the vertical direction. We find that both components are important throughout the solar atmosphere, even when we do not introduce any horizontal field at the bottom boundary to model flux emergence. In figure~\ref{fig:poynting_flux} we show these Poynting fluxes as a function of height and as a function of time at a specific height $z=1.3$~Mm above the photosphere. While both magnetic fields and fluid velocities are possible to measure near the photosphere, it is clear that measuring the Poynting flux \citep[see][]{2015PASJ..tmp..156W} is going to be extremely difficult unless all the terms in equations~\ref{eq:poynting1} and \ref{eq:poynting2} can be accurately estimated. In the chromosphere, transition region and lower corona the two components of the vertical Poynting flux are highly time variable and largely have opposite signs: the work done by horizontal motions results in a largely upward directed Poynting flux, while the transport of horizontal field gives a downward directed flux. However, the time average of the total vertical Poynting flux is directed upward above the photosphere. We find $12\rm~kW~m^{-2}$ at $300\rm~km$, at $1\rm~Mm$ this has fallen by almost an order of magnitude, to $1.5\rm~kW~m^{-2}$, and at 3.5~Mm is further reduced to $500~\rm W~m^{-2}$. Only some $100\rm~W~ m^{-2}$ remains at 10~Mm. The right panel of figure~\ref{fig:poynting_flux} shows that after an initial transient, the total Poynting flux varies on a time scale of roughly 5~minutes and is dominated by  chromospheric and p-mode like oscillations in the atmosphere. During the first 15 minutes the Poynting flux is large, as stresses are built up in the magnetic field, while the Joule dissipation slowly rises as the field gradients approach a critical value at which dissipation is balanced by the buildup of gradients. After this time the total Joule dissipation is remarkably constant and well balanced by the total radiative losses. We shall see that on smaller scales the Joule dissipation is much more variable, but on large scales the system settles fairly rapidly to a quasi steady heating regime. { We note that there is systematic subduction of magnetic field during this run. Thus, on time scales longer than the runs discussed here chromospheric and coronal heating will gradually be reduced unless the magnetic field is replaced, either through a local surface dynamo or through the emergence of new flux from the deeper convection zone.}

\subsection{MODEL B}
 
Model B has a computational box with physical dimensions of  $24\times 24\times 16.8$~Mm in a grid of $768\times768\times768$ points. The grid extends approximately $2.4~\rm Mm$ below $\tau_{500}=1$ and $14.4~\rm Mm$ above. Similar to Model A the grid is uniformly spaced in the $x$ and $y$ direction, about $31~\rm km$ and non-uniform in the $z$ direction with a minimum spacing of $12~\rm km$ in the photosphere up to $82~\rm km$ in the upper corona. The experiment has been run at this resolution for about $30$~minutes solar time, after having been run at lower resolution for an extended period before that. It then takes of order $10$~minutes for the simulation to adjust to the new resolution. We defined $t=0$~s as the instant in which the simulation was interpolated for the present resolution of model B. This model is constructed to have two more or less unipolar magnetic regions separated by some 8~Mm. The average unsigned field strength is 50~Gauss and the net field strength is close to zero. This model is designed to be typical of a quiet sun network region. 

\begin{figure*}
\includegraphics{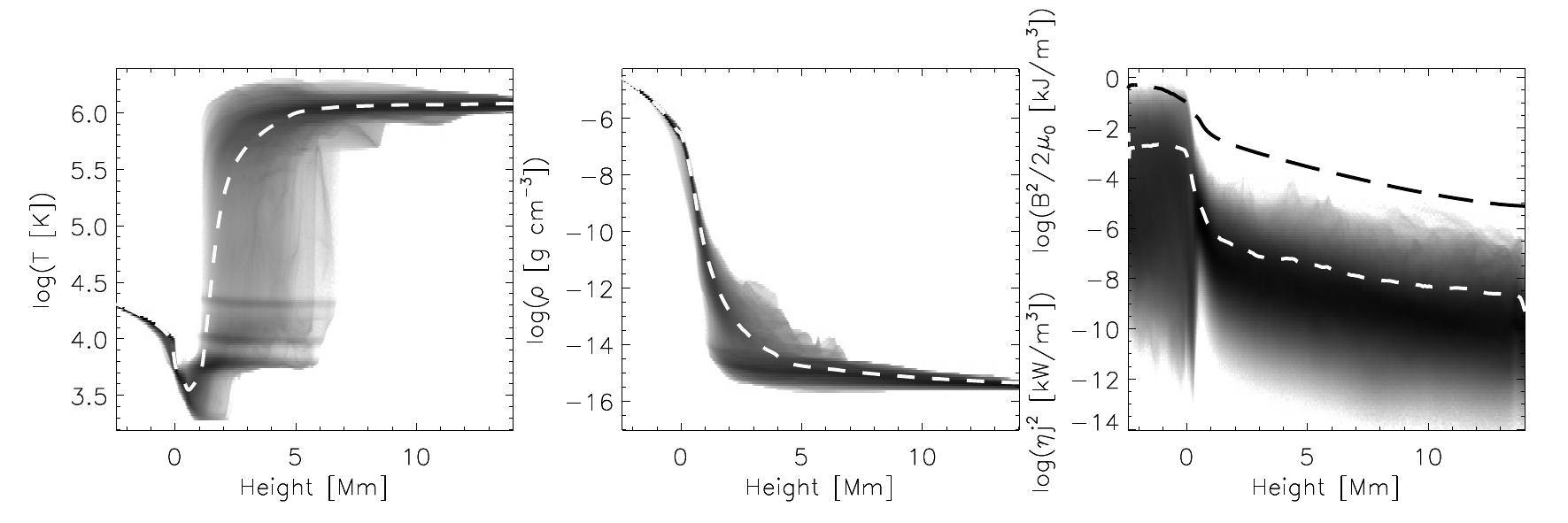}
 \caption{Same as Figure \ref{fig:histem_den_heatlr}, but for Model B.} 
 \label{fig:histem_den_heathr}
\end{figure*}

The distribution of the density, temperature, and heating as a function of height in this model is roughly similar to Model A, as can be seen in figure \ref{fig:histem_den_heathr}. The temperatures in the left panel show a minimum temperature of $2\times 10^{3}$~K and a maximum of $1.6\times 10^{6}$~K. The upper chromosphere and transition region are even more spread out in height than in Model A and we find relatively low temperatures and high densities to cover the height range between $2$~Mm and $6.5$~Mm. The average Joule heating with height also here in Model B has a scale height that is similar to the scale height of the magnetic energy density, and is of the order of $\sim 1600~\rm km$. The Joule heating above $2$~Mm has a scatter ranging roughly 8 orders of magnitude. 
 
\begin{figure*}[!t]
\begin{center}
 \subfigure{\includegraphics[width=.49\textwidth]{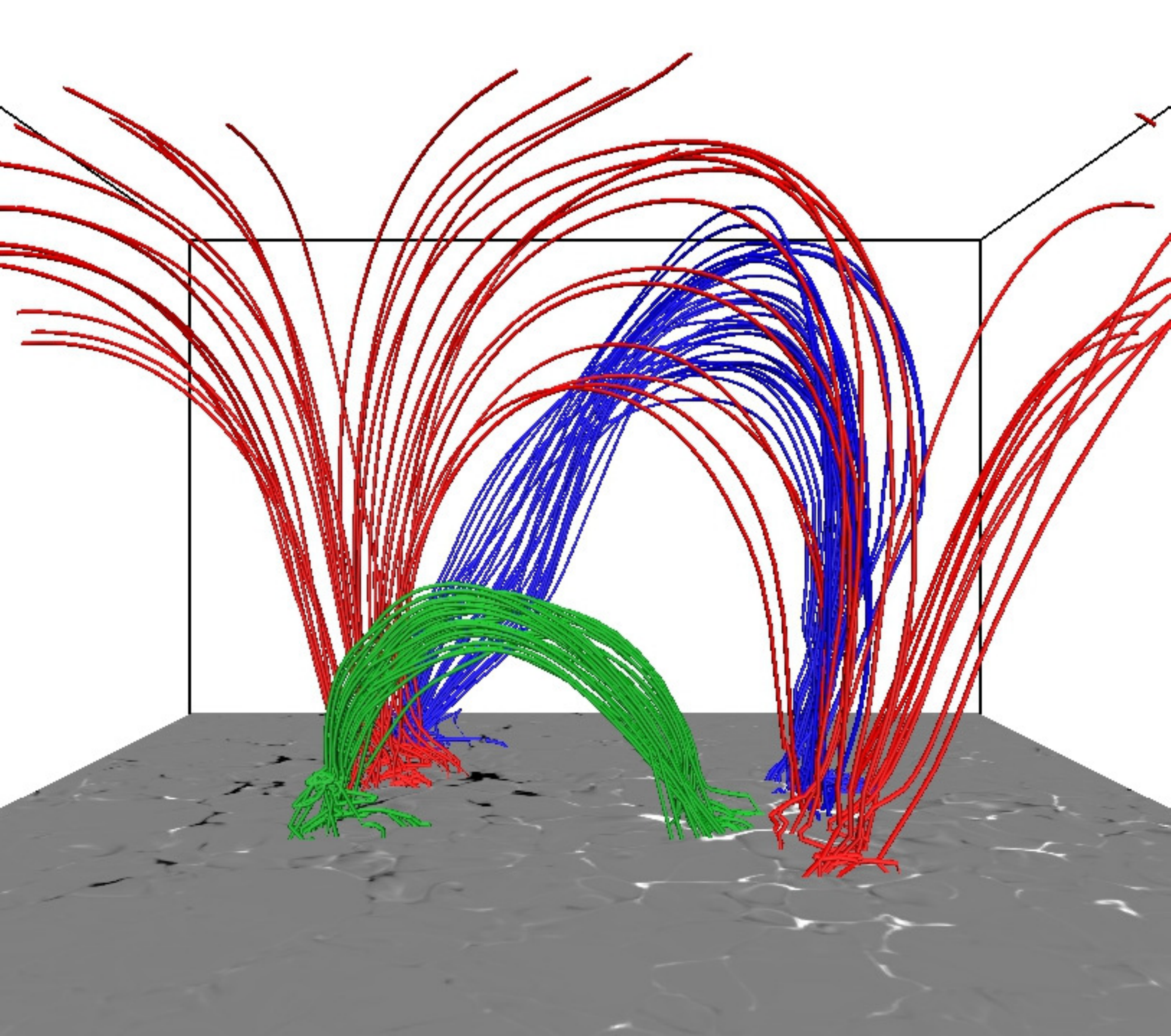}}
  \subfigure{\includegraphics[width=.49\textwidth]{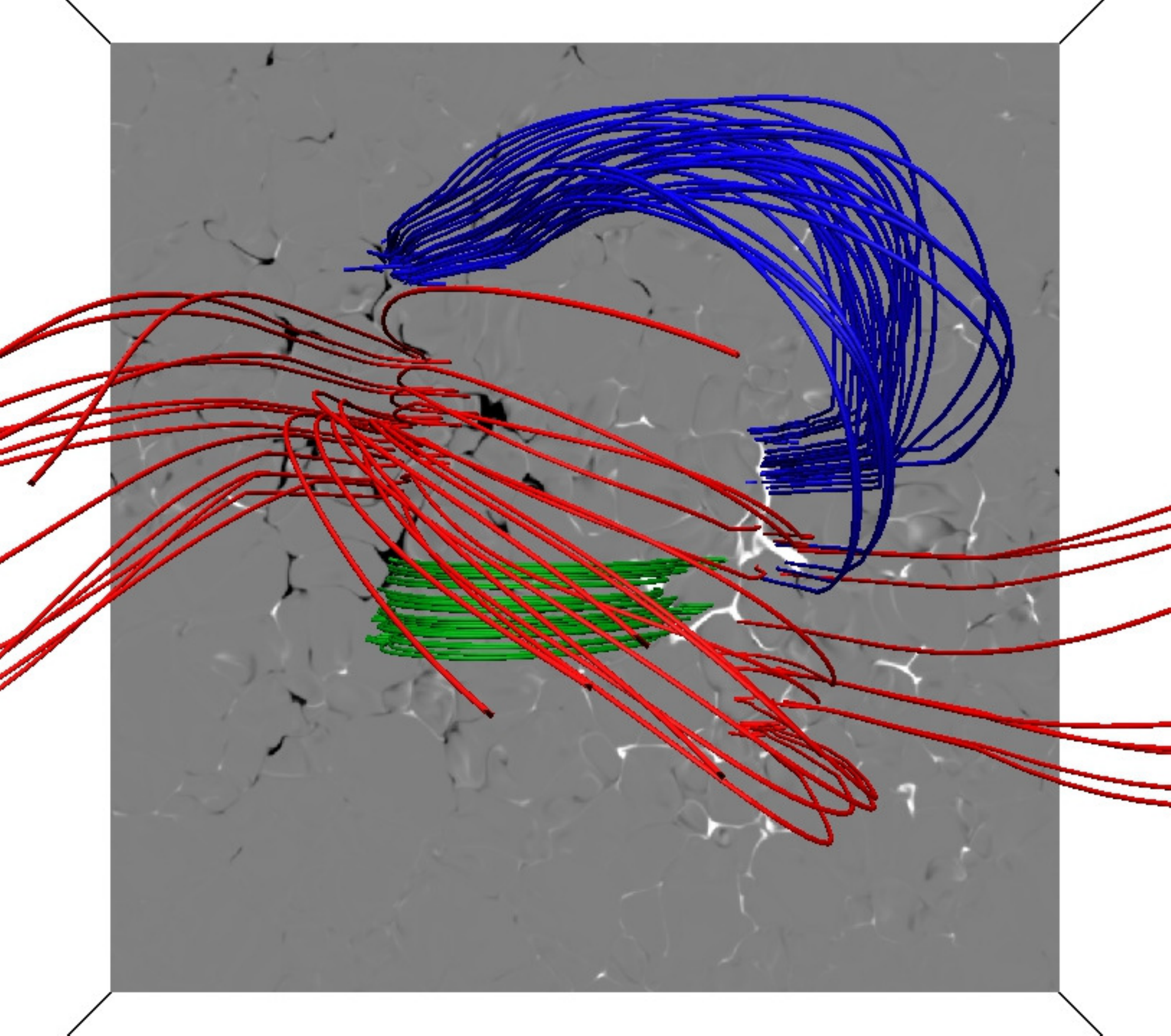}}
  \subfigure{\includegraphics[width=0.49\textwidth]{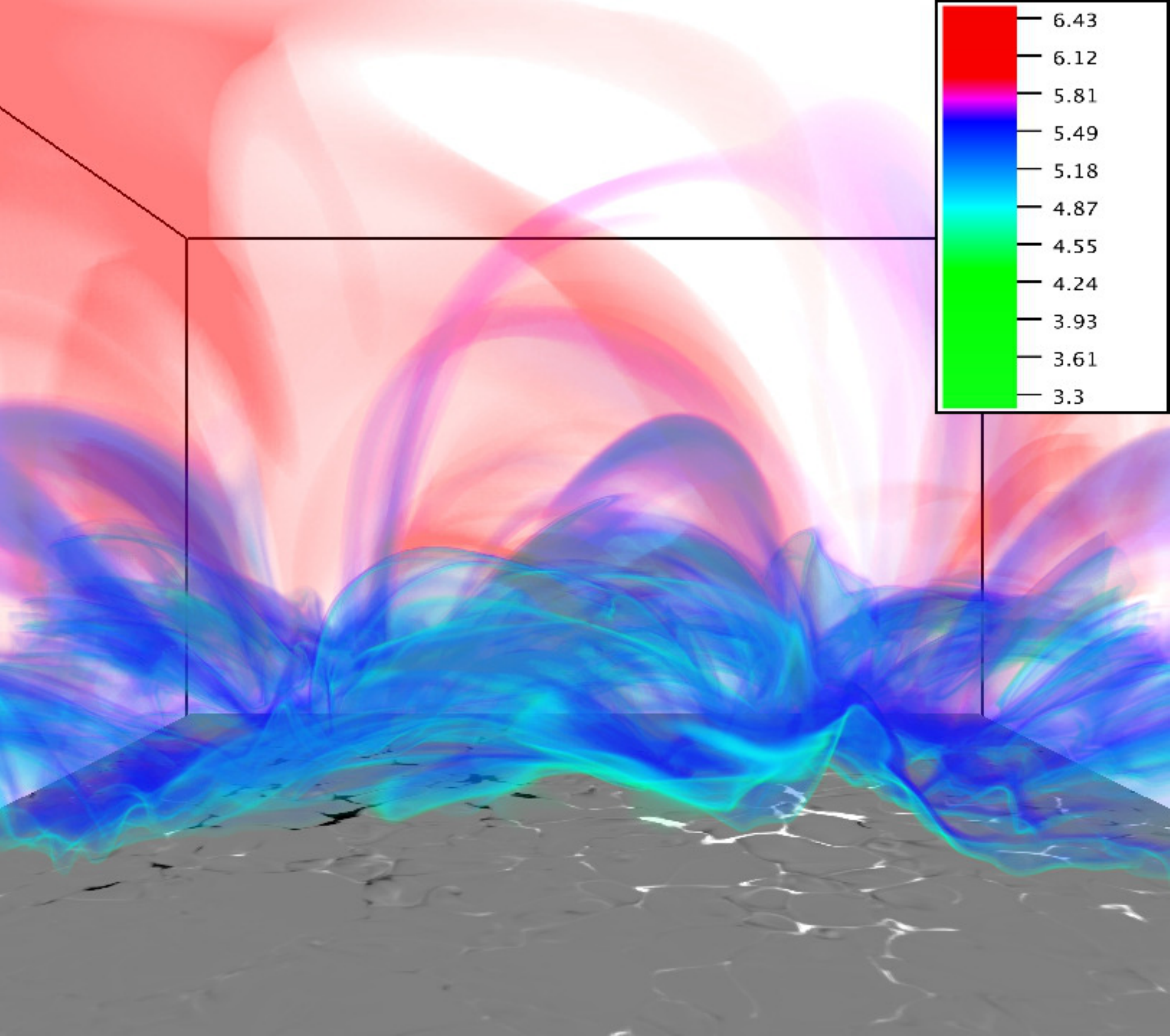}} 
 \subfigure{\includegraphics[width=0.49\textwidth]{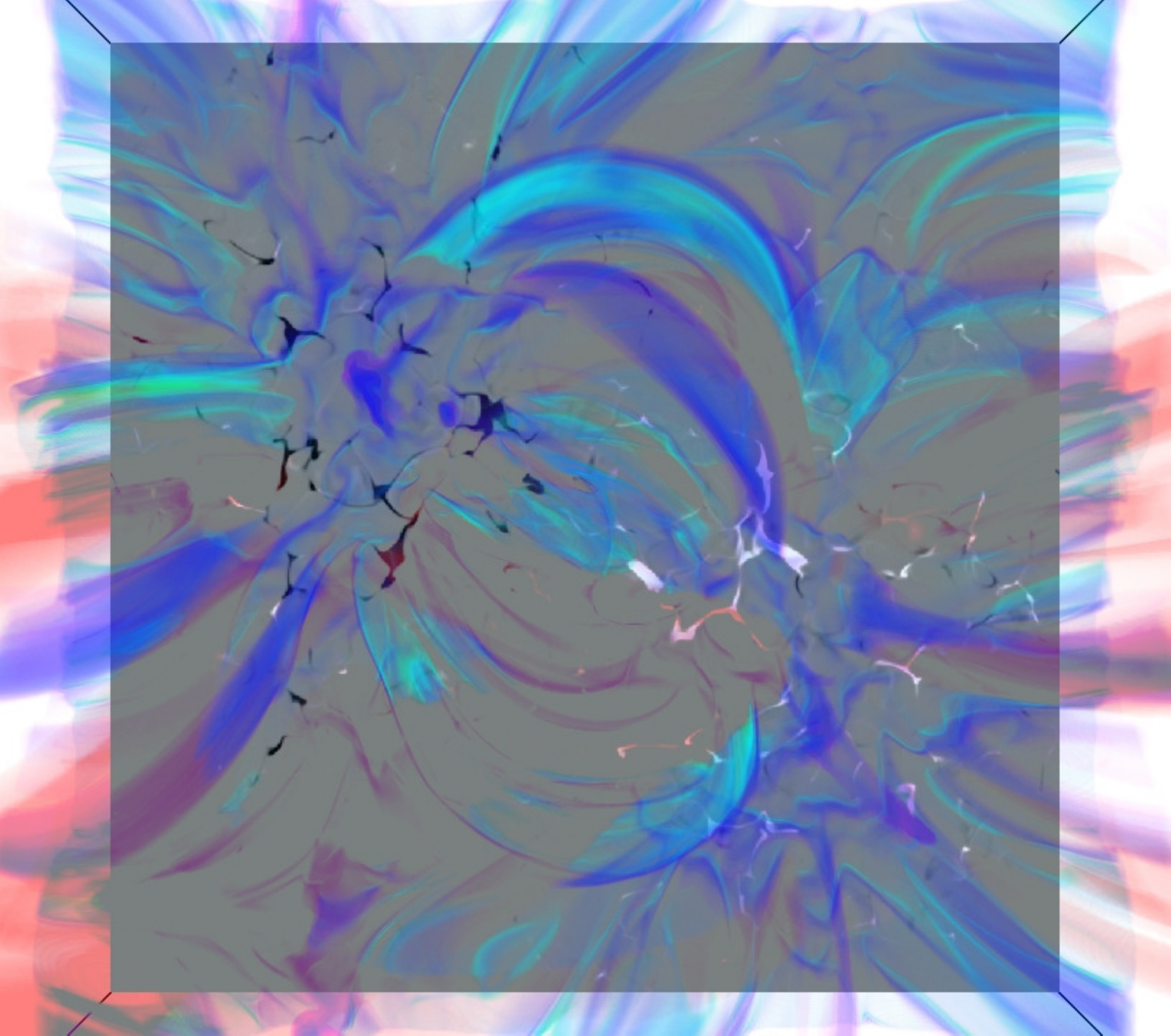}}
 \caption{Magnetic field configuration shown for loops with apex at different heights (top panels: side view-left, top view-right). Temperature stratification (bottom panels: side view-left, top view-right). The temperature color scale is green and turquoise for the chromosphere, blue for the upper transition region, purple for the low coronal temperature and red for the hottest coronal regions. These representations use data from Model B.} 
 \label{fig:MF_modelshr}
\end{center}
\end{figure*}

The bipolar structure of the magnetic field is evident in the photosphere as two regions of concentrated magnetic field of opposite polarity, as shown in figure \ref{fig:MF_modelshr}.  The $B_{z}$ component of the magnetic field in the photosphere at $z = 0$~Mm is shown from the side (left panels) and from the top (right panels). Magnetic field lines are drawn to give an impression of the topology of the stronger parts of the field and are seen to stretch between regions of opposite polarity reaching various heights. This particular figure shows the situation at  $23.8$~minutes after the start of Model B run.

Aspects of the temperature structure are shown in the bottom panels. Upper chromospheric temperatures are largely confined to heights around $2\rm~Mm$, but one also sees isolated cool loops stretching much higher. The same is true for the lower transition region temperatures. Coronal temperatures are found everywhere above $2\rm~Mm$, tending to be found almost universally at the greatest heights, but we find that the hottest isolated loops often are low lying and stretched between regions of the strongest field concentrations. 

\subsection{GLOBAL DISSIPATION PROPERTIES}

As shown in figure~\ref{fig:histem_den_heatlr} and figure~\ref{fig:histem_den_heathr}, in the upper chromosphere and above, the Joule dissipation in these models has a scale height that is roughly the same as that of the magnetic energy, $B^{2}/2\mu_{0}$. We find a scale height of some $650$~km above $z=500$~km in Model A and $z=900$~km in Model B. Below this height, in the photosphere and lower to mid chromosphere, the heating scale height is much smaller, only $100$~km in Model A and $70$~km in Model B.

\begin{figure*}
\includegraphics{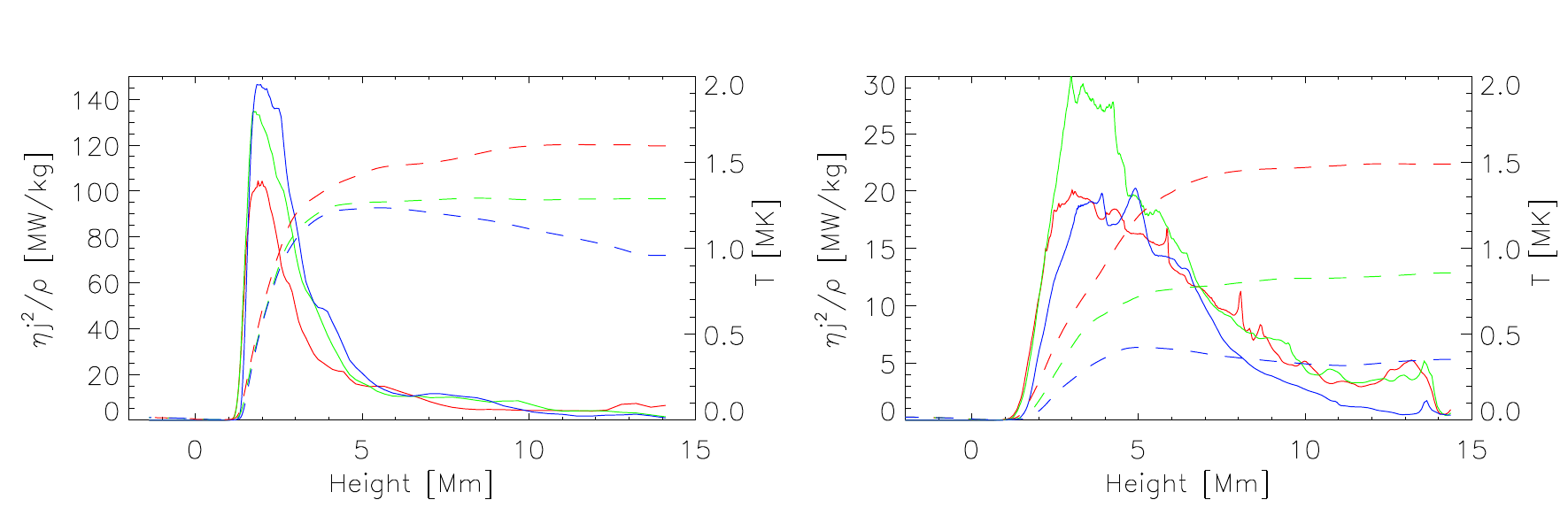}
\caption{Average Joule heating per unit mass $\eta j^{2}/\rho$ [solid line],  and average temperature [dashed
   line] versus height at $t = 1500$~s (blue), $t = 2500$~s (green) and $t = 3600$~s (red) for Model A (left) and at $t = 150$~s (blue), $t = 950$~s (green) and $t = 1750$~s (red) for Model B (right).} 
 \label{fig:j2rt}
\end{figure*}

In the mid to upper chromosphere and above, the heating scale height is larger than the $200\rm~km$ pressure and/or density scale height in the chromosphere, but is much smaller than the coronal scale height of $50\,000$~km. Therefore, as shown by \cite{V.Hansteen082010}, the  Joule dissipation per unit mass ($\eta j^{2}/\rho$), or equivalently per particle, show a maximum in the transition region and low corona: in the chromosphere the number of particles decreases very rapidly and the heating per particle grows exponentially with height, eventually raising temperatures to coronal values. At this point the pressure scale height is much larger than the dissipation scale height and the heating per particle decreases, also exponentially, but with the longer scale height of the magnetic energy density. This is shown in figure \ref{fig:j2rt} along with the average temperature, as a function of height, showing the situation at three different times for both Model A and Model B.

The maximum average Joule dissipation per unit mass is about one order of magnitude higher for Model A than for Model B, presumably due to the stronger average field in Model A (a factor $\times 3$), but perhaps also due to the different field topologies of the models. The decrease of the heating per particle with height is less steep for Model B than for Model A.  The magnetic field scale height for Model A is smaller than the magnetic field scale height for Model B above the height of maximum Joule dissipation, which is placed above $2$~Mm in both models. 
Note also that the average temperature rise in Model B is less steep than in Model A, a consequence of the greater amount of cool material at great height in Model B simulation.

\begin{figure*}[!t]
\begin{center}
 \subfigure{\includegraphics[width=.49\textwidth]{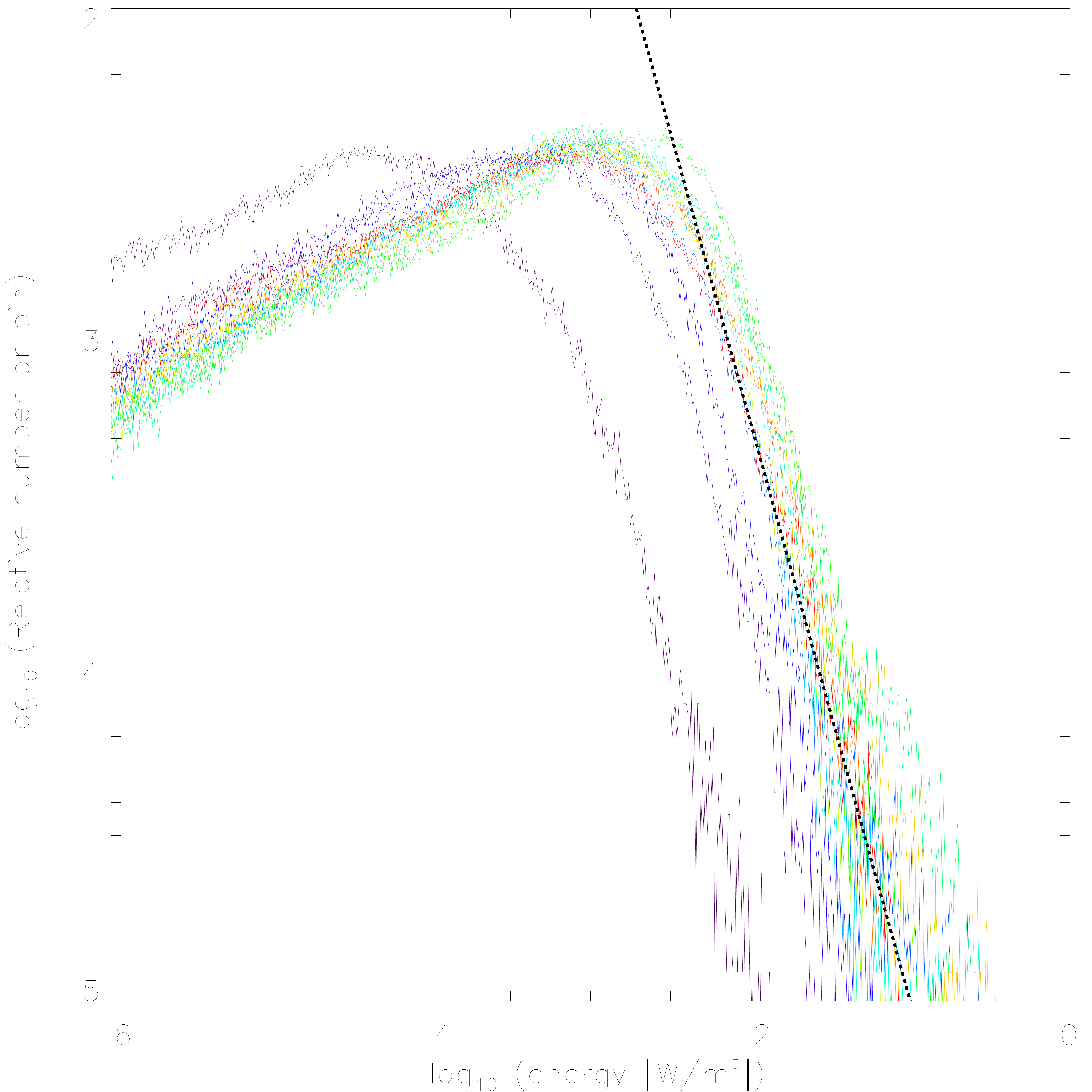}}
 \subfigure{\includegraphics[width=.49\textwidth]{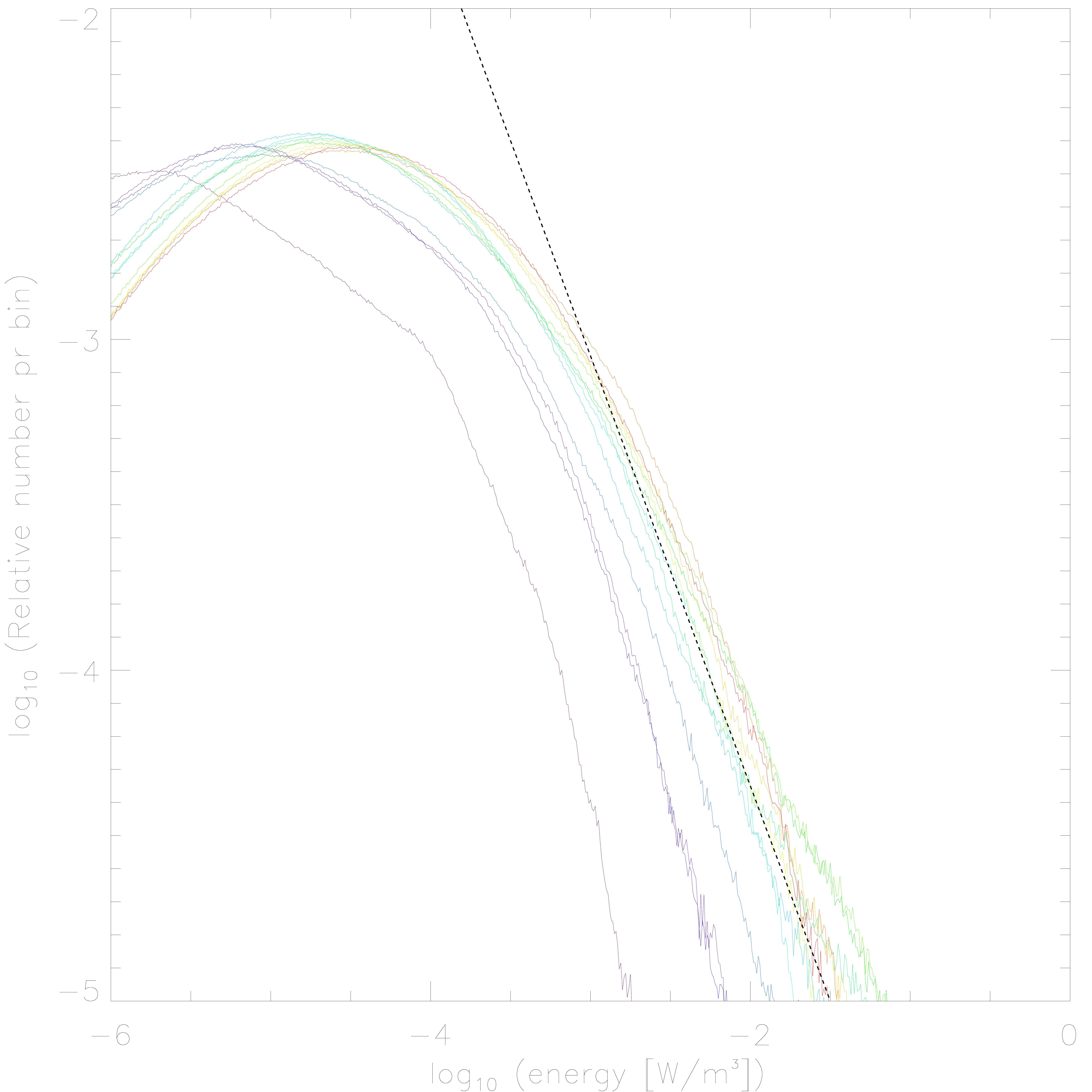}} 
  \caption{Histograms of the Joule heating in a region roughly at  $2$~Mm height for Model A (left) and Model B (right) using a bin size of 0.01 in the log. The dotted lines have slope equal to $1.75$ for Model A and $1.3$ for Model B and are adjusted by eye. The colors represent the times of the different snapshots: from purple in the beginning of the run to red at the end. In Model A we plot the distribution at $t=10$~s, $100$~s, $150$~s, $200$~s, $300$~s, $400$~s, $500$~s, $1000$~s, and thereafter every $500$~s until $3500$~s, and at $t=10$~s, $50$~s, $100$~s, $150$~s, $200$~s, $600$~s, and thereafter every $200$~s until  $2000$~s for Model B.} 
 \label{fig:histograms}
\end{center}
\end{figure*}

As should be obvious from figures~\ref{fig:histem_den_heatlr} and \ref{fig:histem_den_heathr} the Joule heating is not all found in events of a certain magnitude, events are rather spread in energy over several decades. It is difficult to isolate events in space and time, since events move and overlap, but we can make a distribution function of heating magnitudes at a given height. 
In figure~\ref{fig:histograms} we plot histograms of the magnitude of Joule heating in individual mesh points in a $200\rm~km$ high zone centered on $z=2$~Mm at various times for Model A and B using a bin size of 0.01 dex in energy. Initially, both models show very few grid cells with large heating but within a few hundred seconds a more or less constant distribution is established with a given slope. Since there are a finite number of computational cells the distribution is necessarily truncated at low energies and a peak appears; at $\simeq 10^{-3}\rm~W~m^{-3}$ for the Model A, and at $3\times 10^{-5}\rm~W~m^{-3}$ for model B. However, the slopes should be taken with a grain of salt since the chosen bin size is an important parameter in the resulting slope: increasing the bin size would make a steeper distribution while a smaller bin size would result in flatter distributions. There is no physical argument to choose one bin size over another and the index of the power law is not a 
 reliable measure of the role of low energy versus high energy events in the corona. On the other hand figure~\ref{fig:histograms} does show that these models establish a steady state heating within a thousand seconds with a given spread in energetic and less energetic heating.

\section{JOULE DISSIPATION CHARACTERIZATION}
\subsection{HEATING EVENTS}

On the largest, global, scale we find that the Joule heating in the model is (very) roughly proportional to the magnetic energy density. With the magnetic field topologies considered in this paper this implies that the heating is vertically stratified with a scale height of some thousands of kilometers, and that the heating per particle therefore is concentrated in the upper chromosphere, transition region, and lower corona as described in the previous section. Let us now consider the heating on smaller and horizontal scales. As the (originally potential) magnetic field is stressed by convective and granular motions, gradients build up. These gradients rapidly lead to the formation of current sheets as described by \citet{K.Galsgaard061996} along which dissipation and hence heating occur. 

\begin{figure*}[!t]
\begin{center}
 \subfigure{\includegraphics[width=0.49\textwidth]{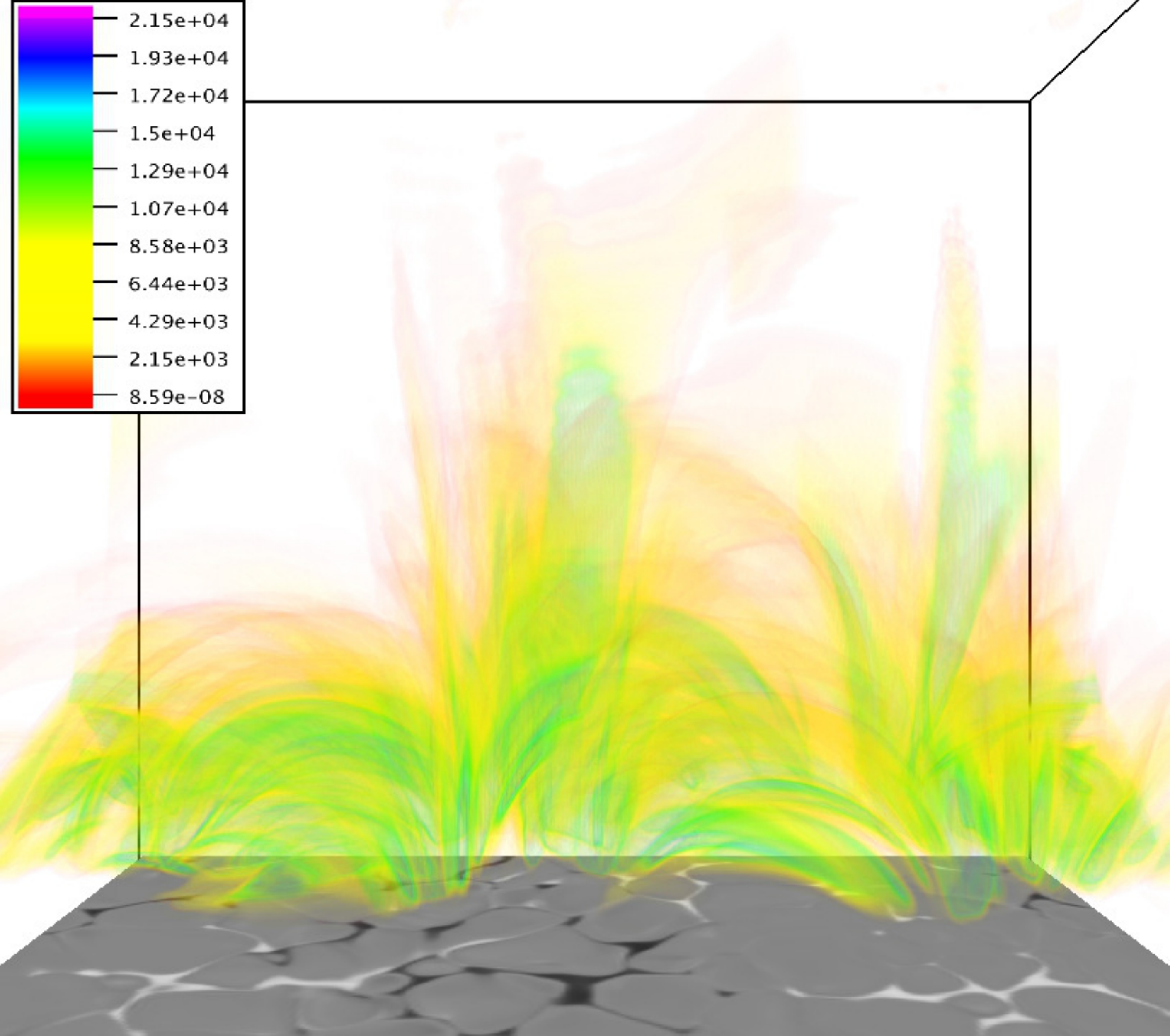}}
  \subfigure{\includegraphics[width=0.49\textwidth]{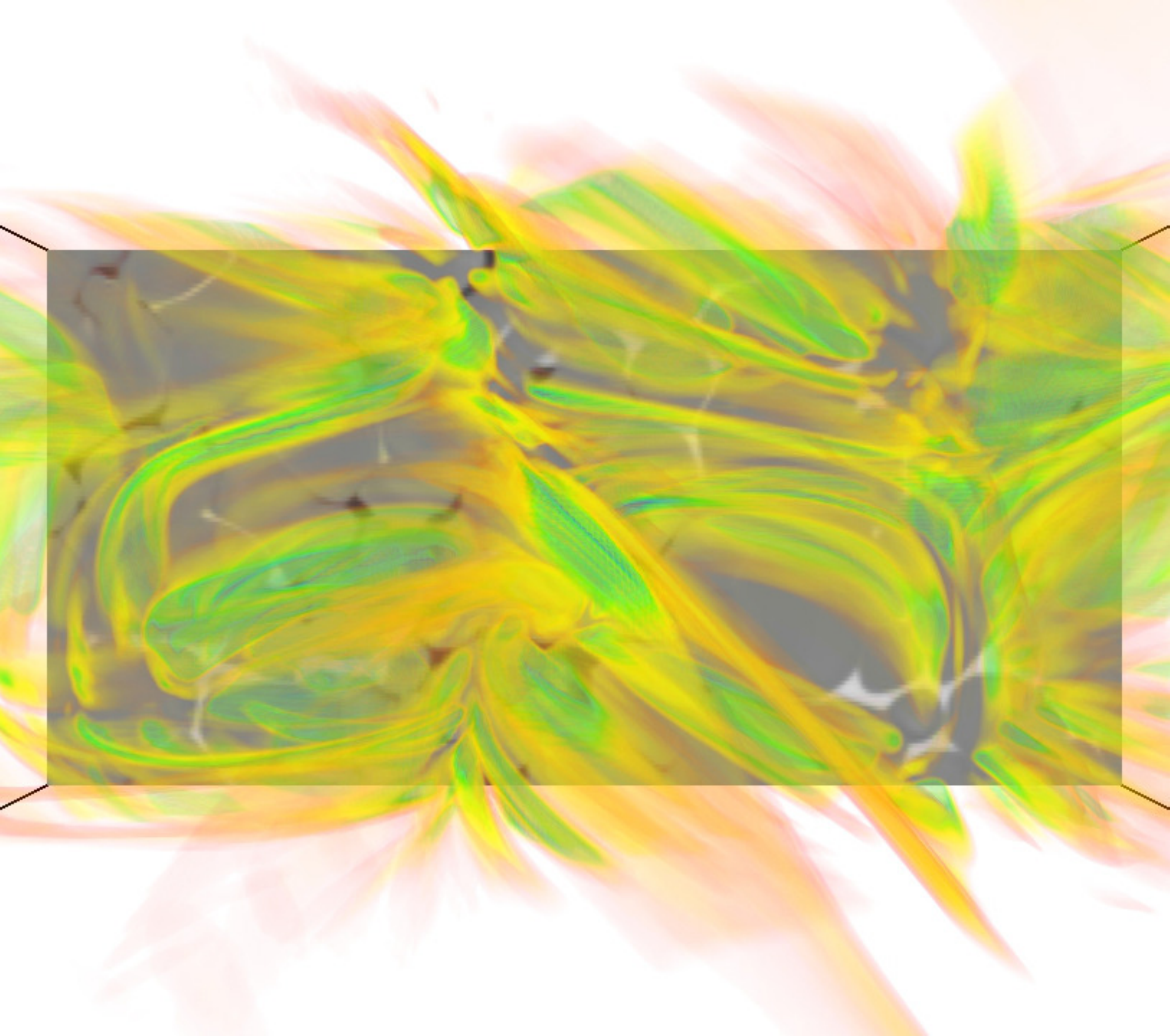}} 
   \subfigure{\includegraphics[width=0.49\textwidth]{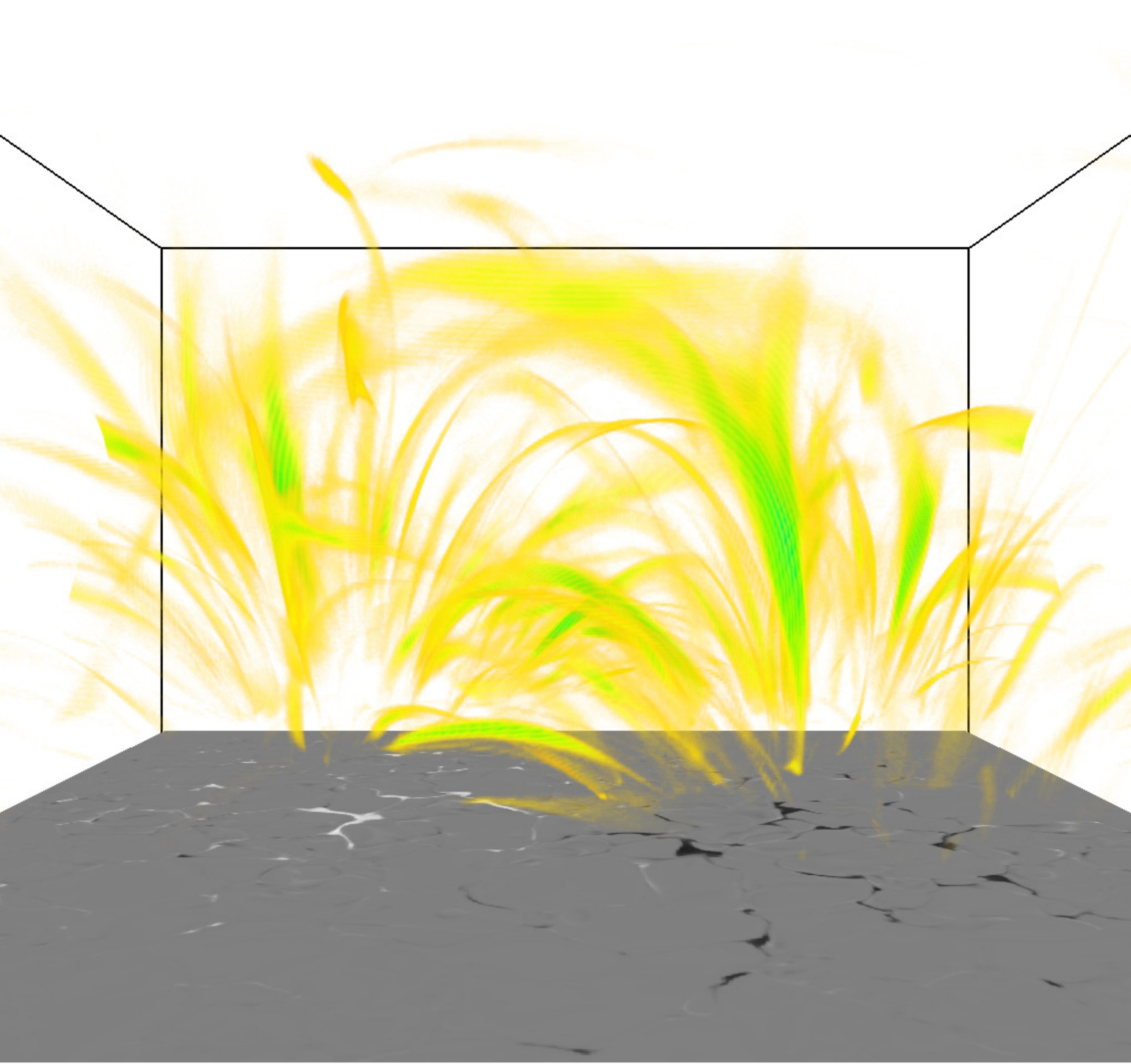}}
  \subfigure{\includegraphics[width=0.49\textwidth]{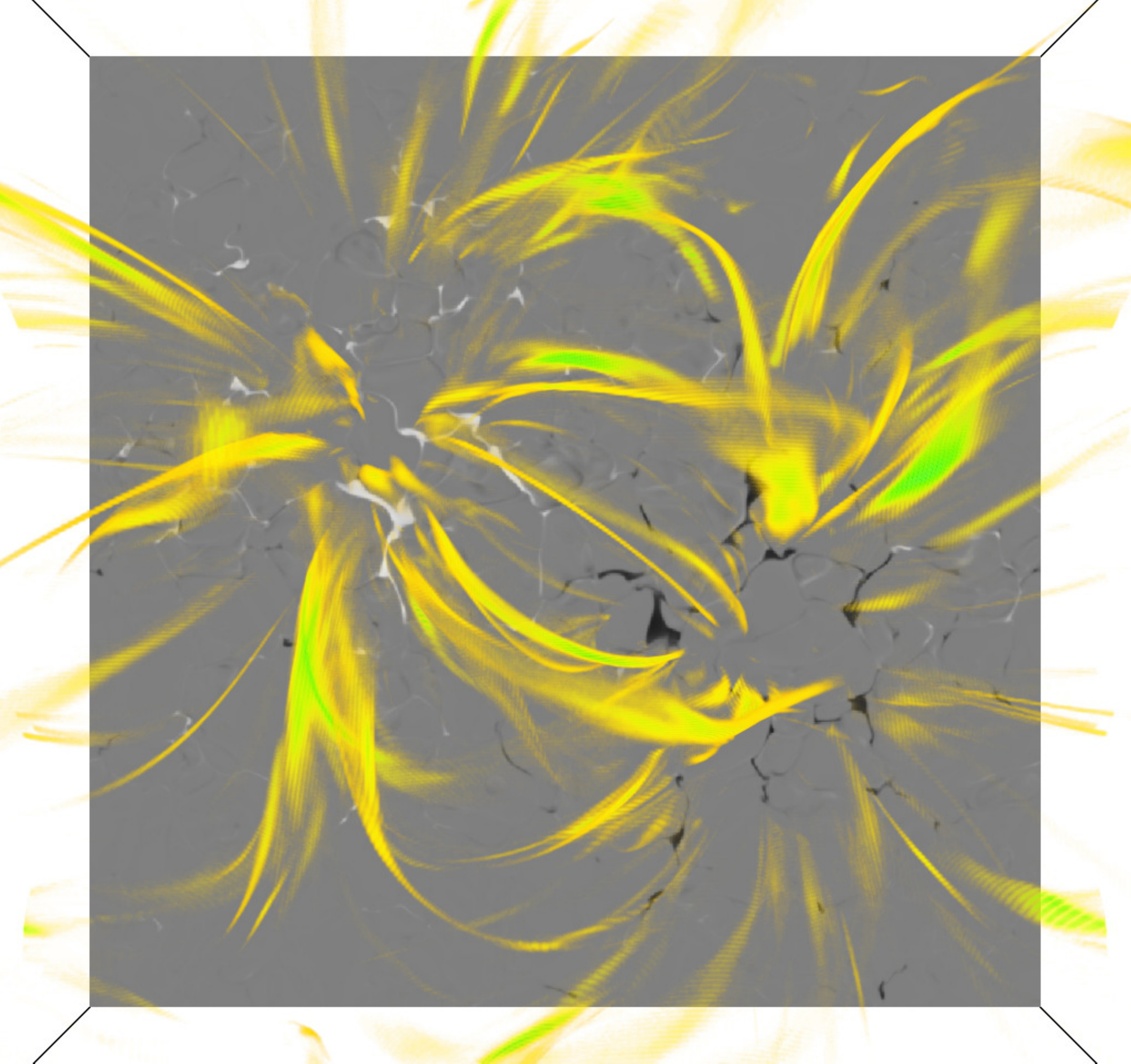}} 
 \caption{Current density squared per particle $\left( j^{2}/\rho\right)$ (arbitrary units), for Model A (top) and Model B (bottom)
[side view-left, top view-right].} 
 \label{fig:JH_Model A_Model B}
\end{center}
\end{figure*}

The spatial distribution of the current sheets is quite complex, as can be seen in figure~\ref{fig:JH_Model A_Model B} where the current density per particle in both models is shown. Note that though quite intermittent in both models, the current sheets generally follow the global structure of the magnetic field as field lines loop from one photospheric polarity to the other. The complexity of the photospheric field in Model A is somewhat greater than Model B: positive and negative polarities are quite intermixed in Model A and there is also a larger net signed magnetic field so that a significant number of field lines escape the top boundary of Model A. The highly intermixed polarities imply a smaller magnetic energy density scale height, and indeed we find that the strong currents sheets in this model are concentrated at low heights outlining loop shaped structures. In addition, we find regions of strong currents that extend significantly higher than the transition region along the
  ``open'' field lines that pierce the top boundary. Model B has a smaller unsigned flux and the opposite magnetic field polarities in the photosphere are better separated. Thus, we find a larger magnetic field energy density scale height and fewer ``open''  field lines: regions of strong current sheets extend higher in Model B, but we do not find ``fingers'' of dissipation extending towards the upper boundary. It also appears that strong current sheets are more spatially intermittent in Model B.

While current sheets stretch out, forming arc or finger shaped structures along the field, they rapidly collapse to dimensions of only a few grid zones perpendicular to the field. This latter scaling is ensured by the functional form of the resistivity which is constructed proportional to the grid size $\Delta s$, resulting in the grid Reynolds numbers of order slightly greater than one that are required in order to resolve the current sheets on the chosen grid. In the two simulations described here this gives a current sheet thickness of some $120\rm~km$ for Model A and some $80\rm~km$ for Model B. Of course, on the Sun, the current sheet thickness in the corona is expected to be much, much smaller, perhaps on the order of centimeters or less depending on the (unknown) physical process that ultimately halts current sheet collapse. Nevertheless, as has been pointed out by numerous authors \citep{K.Galsgaard061996,A.Nordlund001997,B.Gudiksen012005,S.Bingert062011},
the total dissipation in  a given current sheet is independent of the resolution $\Delta s$. On the other hand, while the total dissipation remains the same, the scale at which the dissipation takes place will change with resolution, and this change may have an impact on the local observable consequences of the energy dissipation.

\begin{figure*}[!t]
\begin{center}
\subfigure{\includegraphics[width=.43\textwidth]{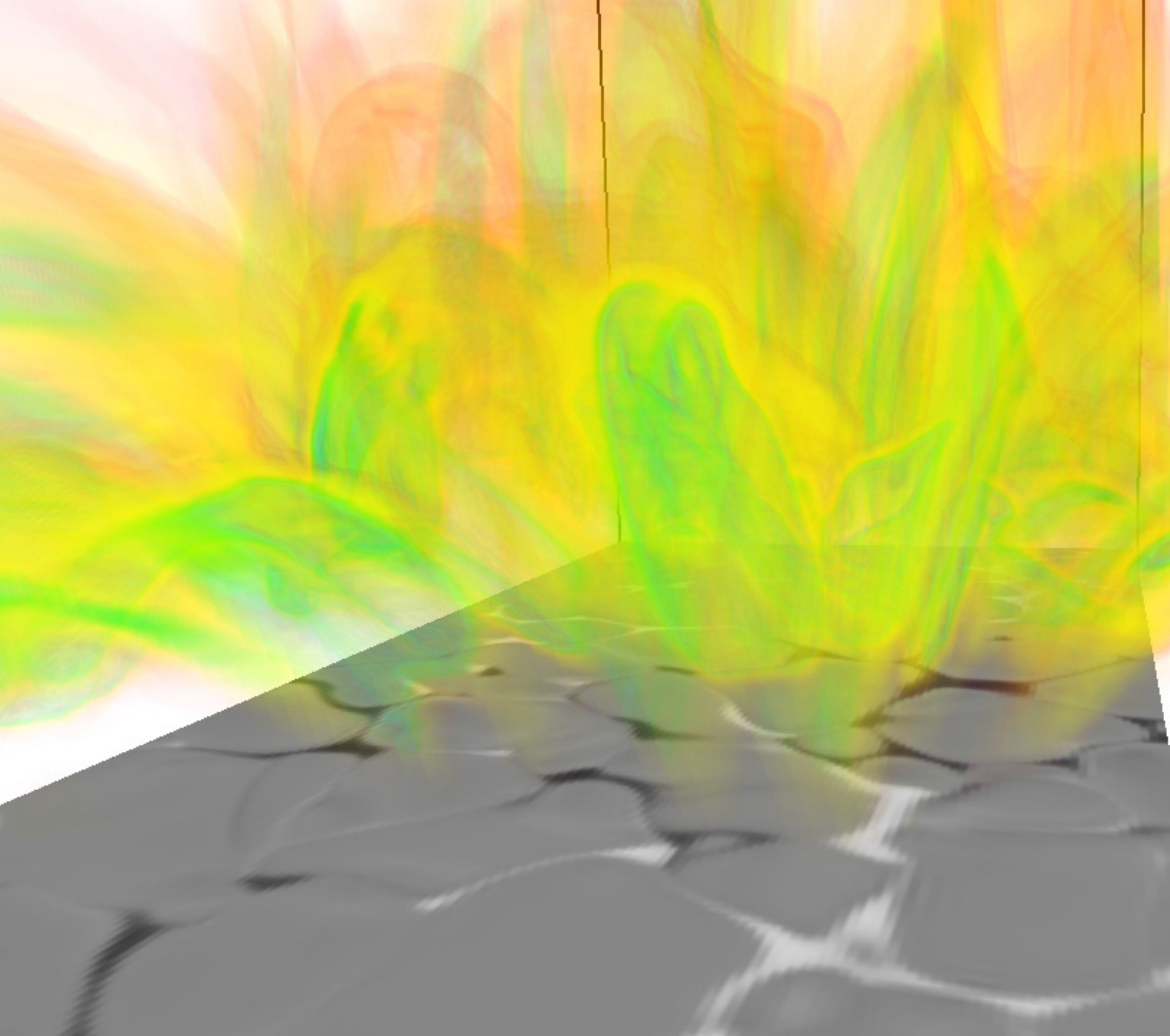}}
 \subfigure{\includegraphics[width=.43\textwidth]{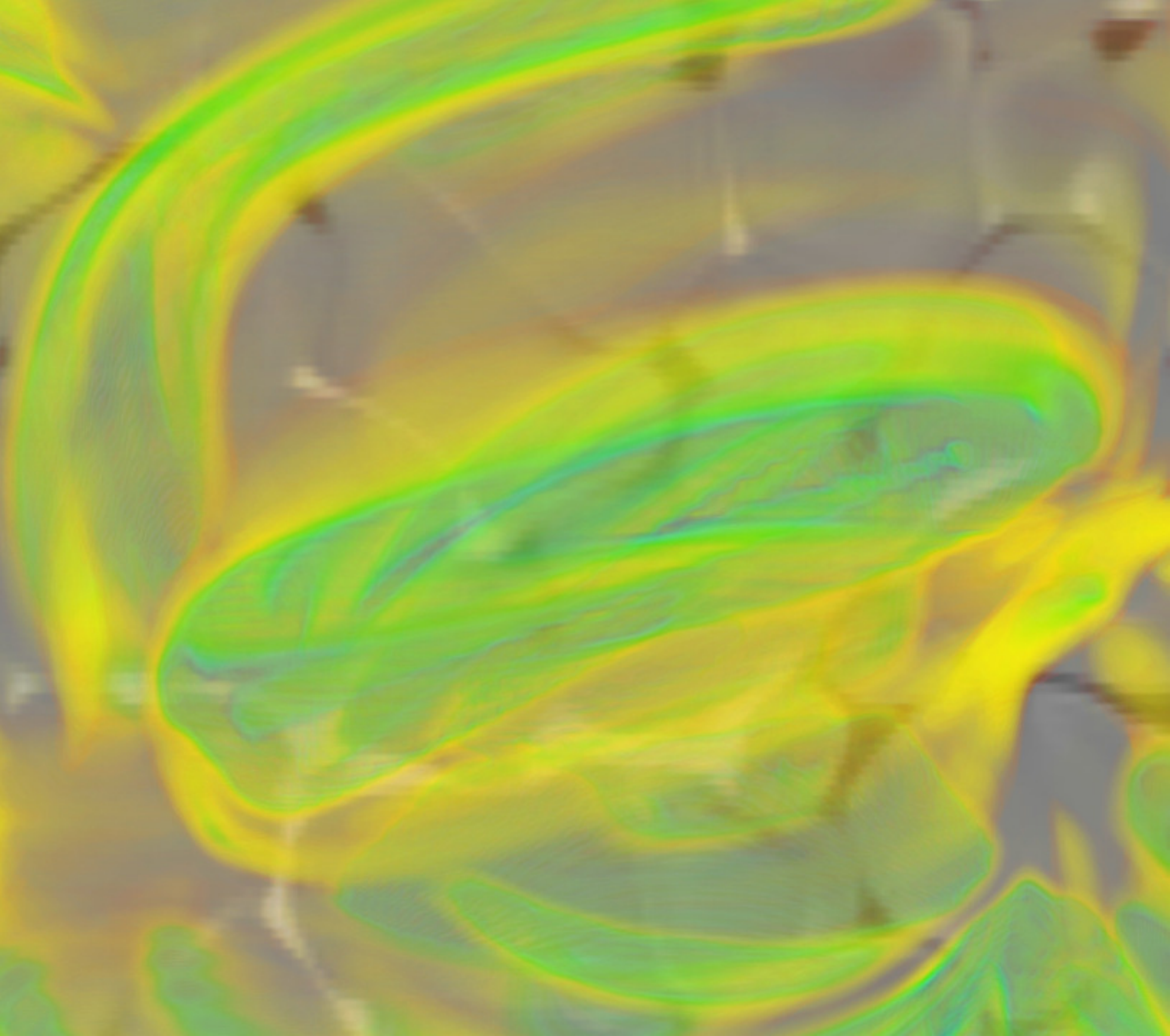}}
  \subfigure{\includegraphics[width=.43\textwidth]{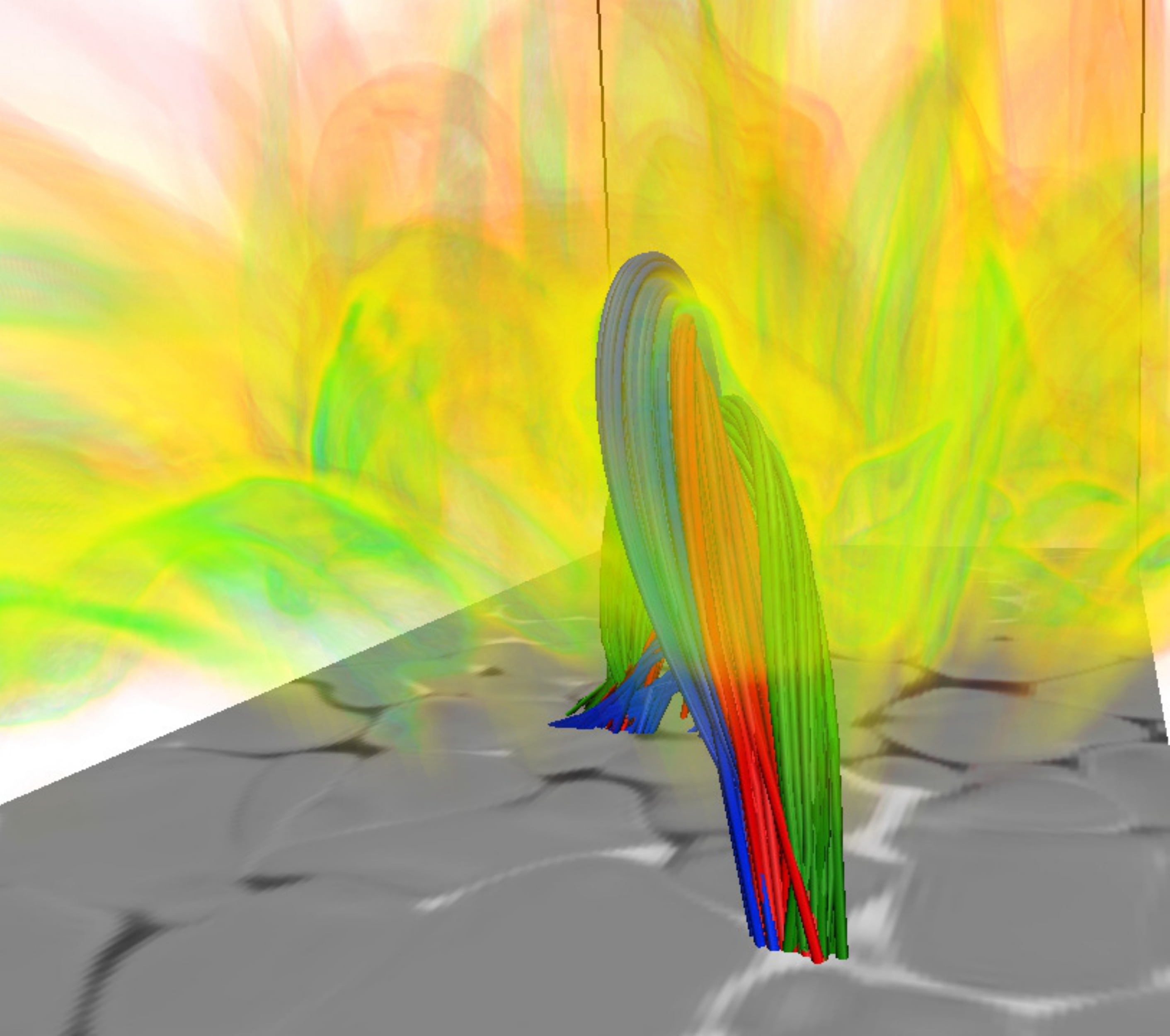}}
 \subfigure{\includegraphics[width=.43\textwidth]{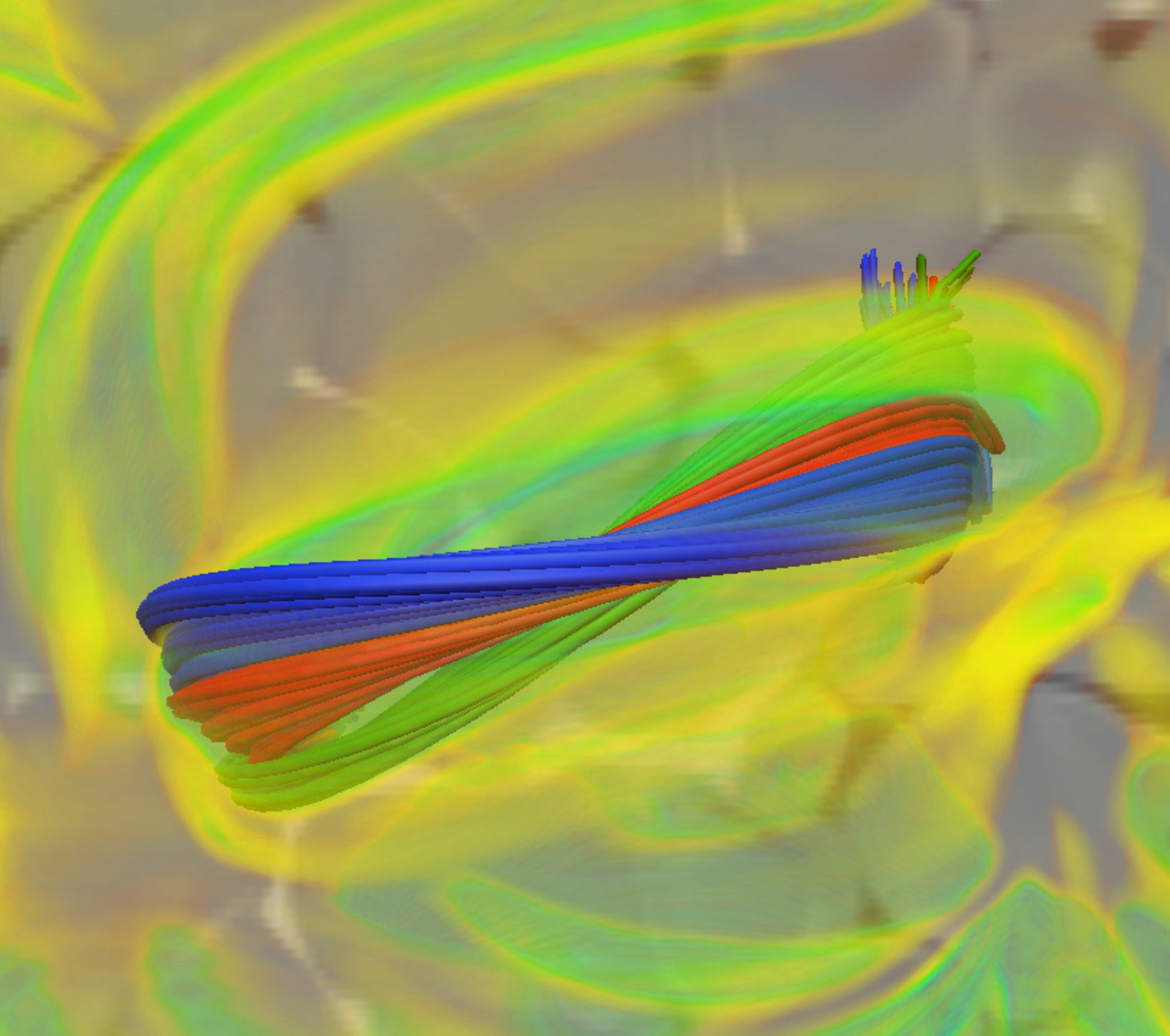}} 
 \caption{Topology of the magnetic field in a region of strong Joule heating per particle at  $t=500$~s for Model A; [left - side view, right-top view] and current density squared per particle ($\left( j^{2}/\rho\right)$), the color table is the same as in Figure \ref{fig:JH_Model A_Model B}. The magnetic field loops have three colors to distinguish the loops above (blue), very close to (red), and below (green) the dissipation region. The right panels have a $4.4\times 3.9\rm~Mm$ field of view.} 
 \label{fig:MF_angle1a}
\end{center}
\end{figure*}
Following \cite{G.Baumann062013}, the current density in the current sheet is given by 

\begin{equation}
j_{\bf C}=\nabla\times{\bf B}\sim\frac{\Delta B}{\Delta s}\sim\frac{\sin \phi  B_{\rm CS}}{\Delta s}
\end{equation}
\noindent
where $\phi$ is the angle characterizing the difference of direction of field lines on either side of the current sheet, set by photospheric motions, and $B_{\rm CS}$ is a typical field strength just outside the current sheet. For the magnetic topologies considered in this paper and in the absence of flux emergence we typically find small angles $\phi$ in the current sheets that dissipate the stresses passed to the outer atmosphere from photospheric motions. A typical example of this is shown in figure~\ref{fig:MF_angle1a} which shows details of a dissipating current sheet at $t=500$~s in Model A. In the upper two panels we show the current sheet as seen from the side and from above --- note that the current sheet shows some internal structure as it is beginning to fragment. Selected field lines passing above and below the current sheet are drawn showing the (small) angle $\phi$ that induces the current. { We note that the small angles found here for the onset of heating episodes (and later, for Model B) are not consistent with the ``secondary instability'' (SI) advocated by \cite[{\it e.g.}][]{2005ApJ...622.1191D} which appear to require substantially larger shear. }

\begin{figure*}[!t]
\begin{center}
 \subfigure{\includegraphics[width=.30\textwidth]{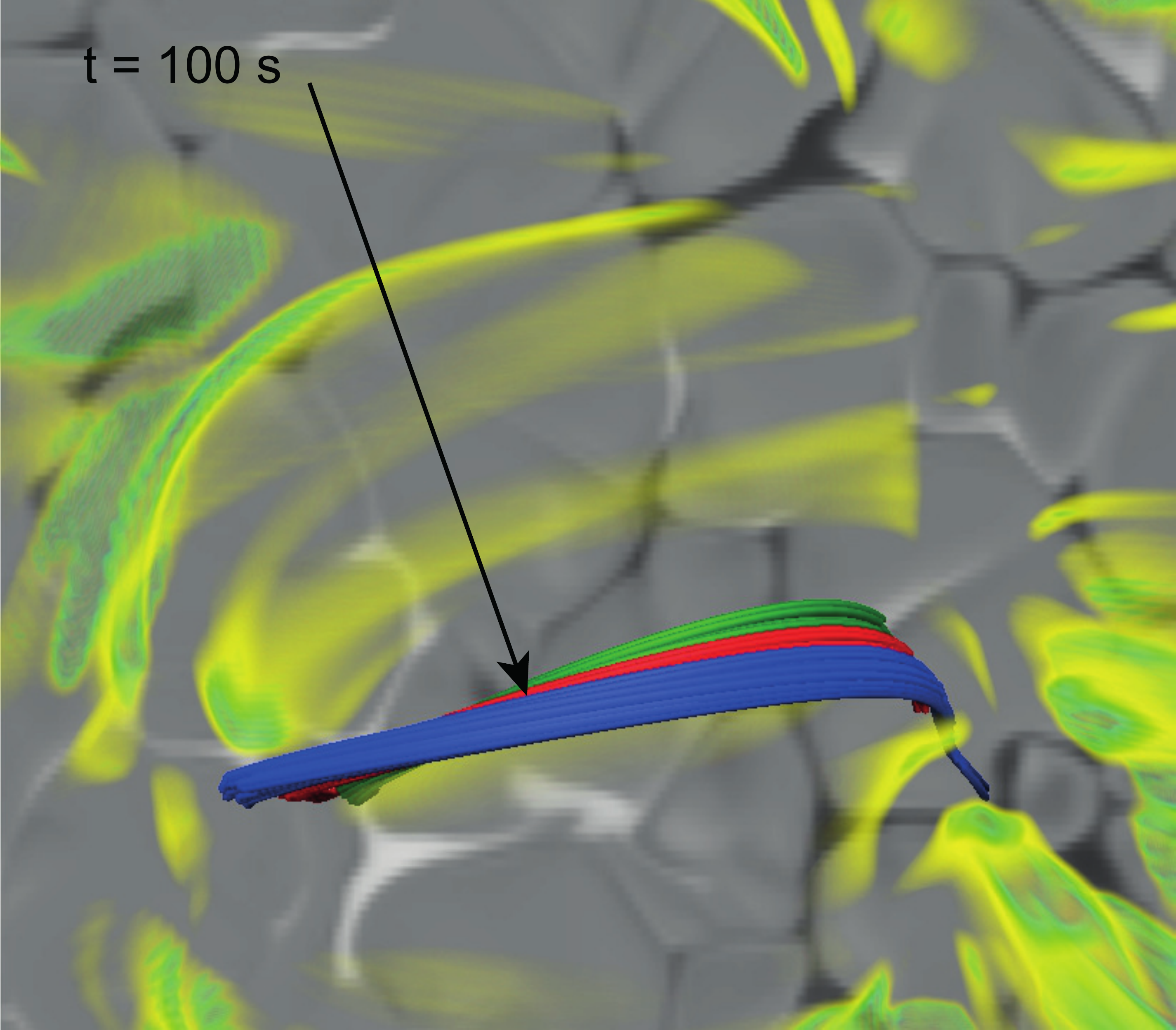}}
 \subfigure{\includegraphics[width=.30\textwidth]{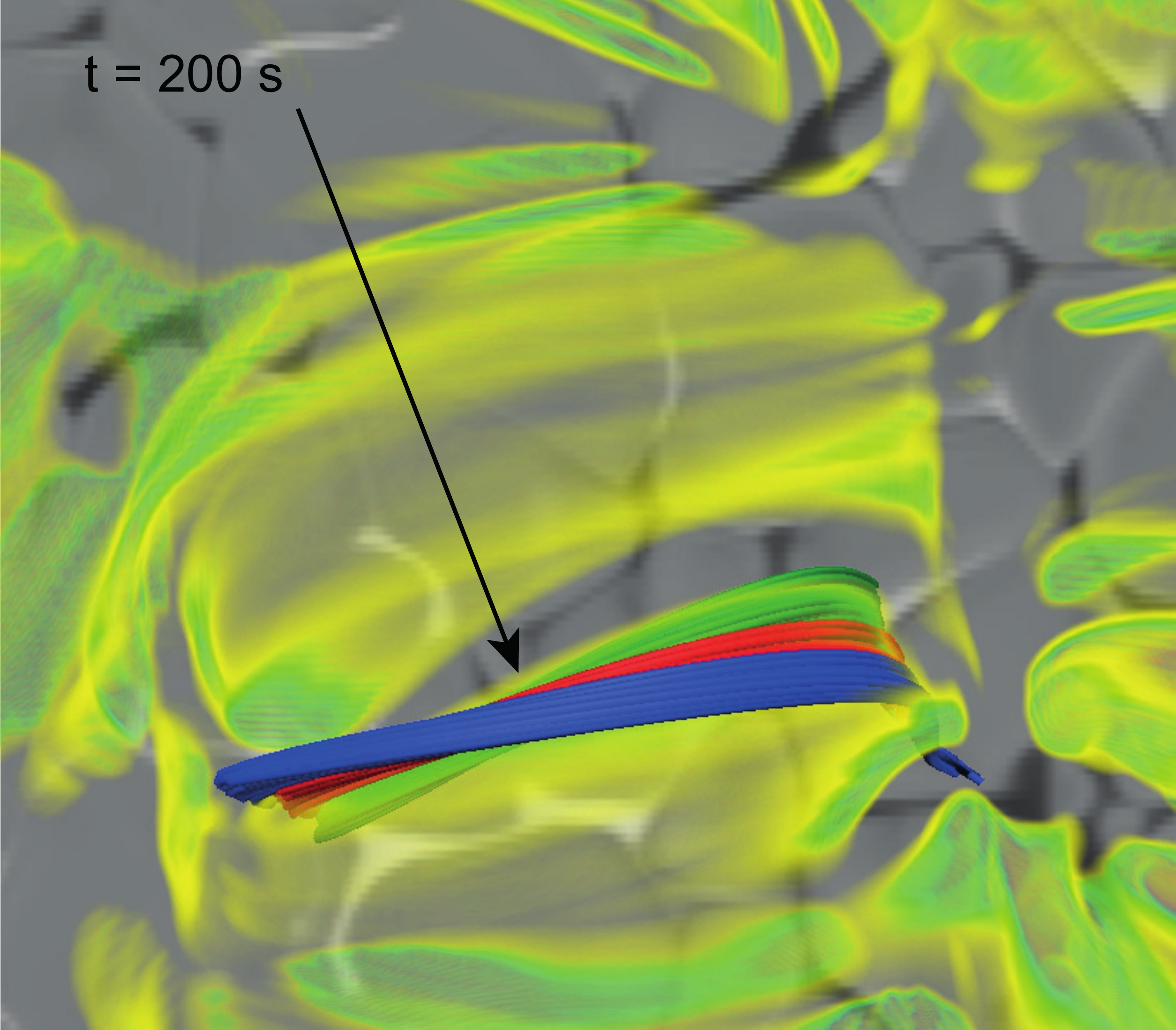}}
  \subfigure{\includegraphics[width=.30\textwidth]{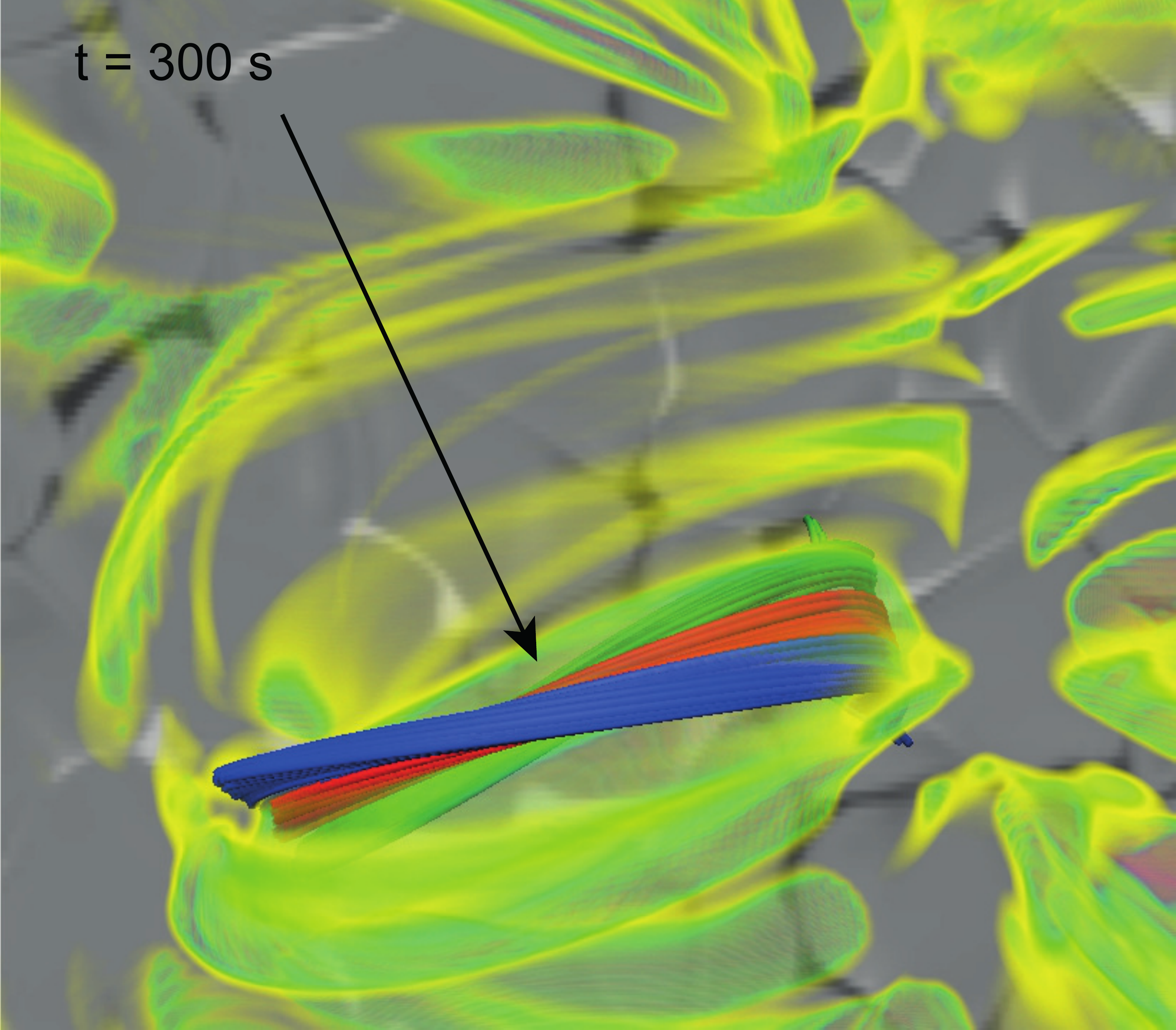}}
 \subfigure{\includegraphics[width=.30\textwidth]{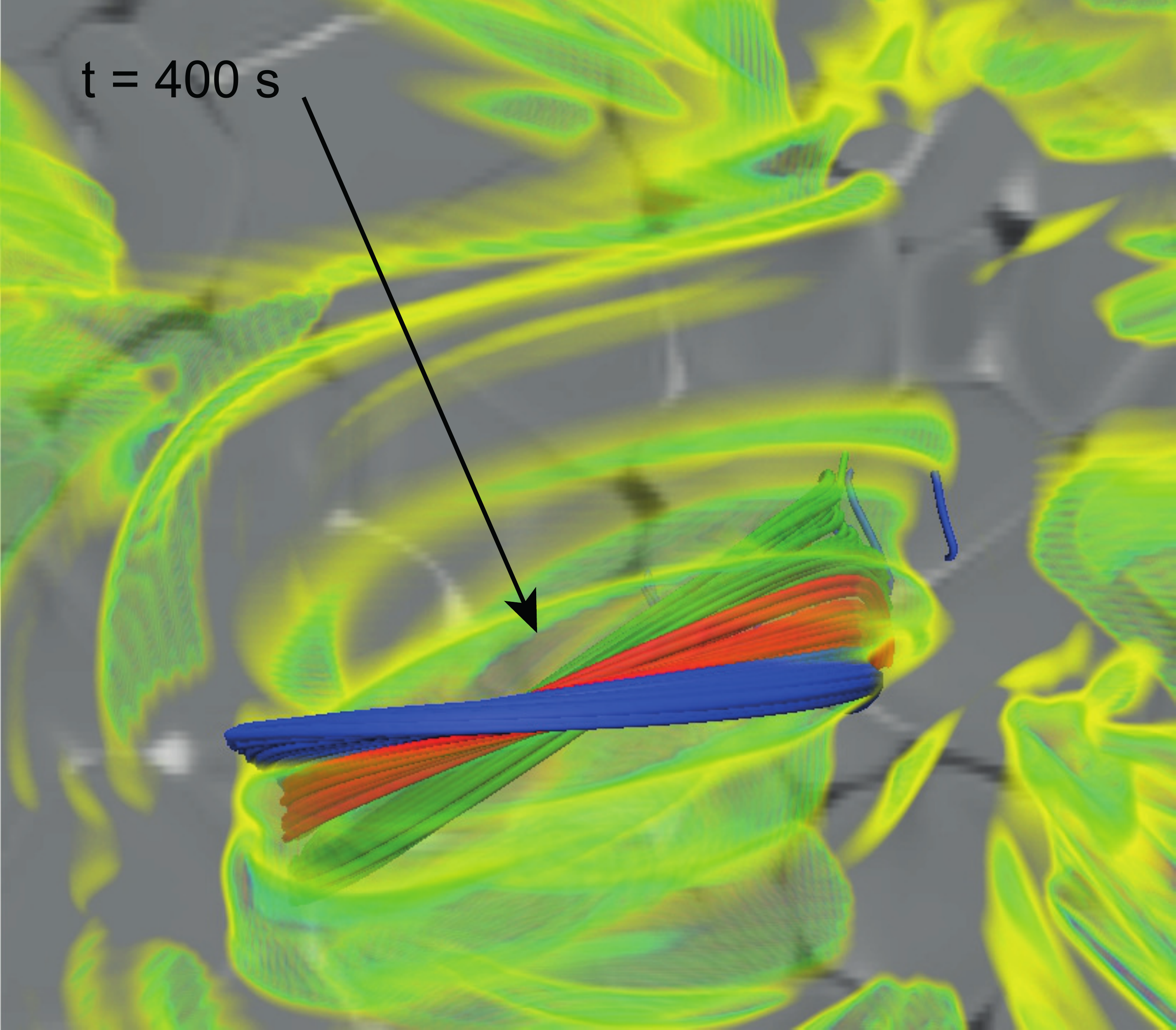}} 
  \subfigure{\includegraphics[width=.30\textwidth]{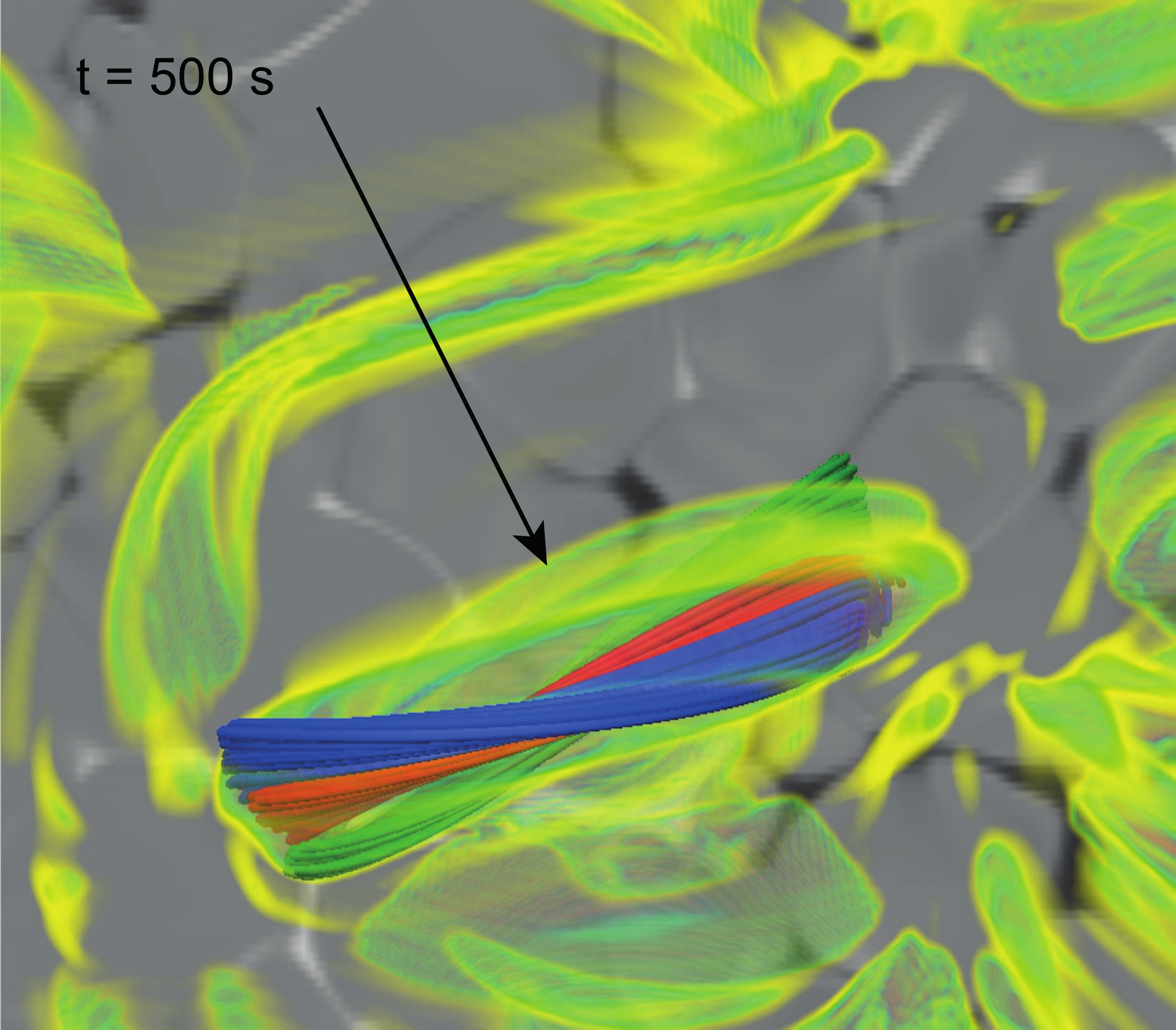}}
  \subfigure{\includegraphics[width=.30\textwidth]{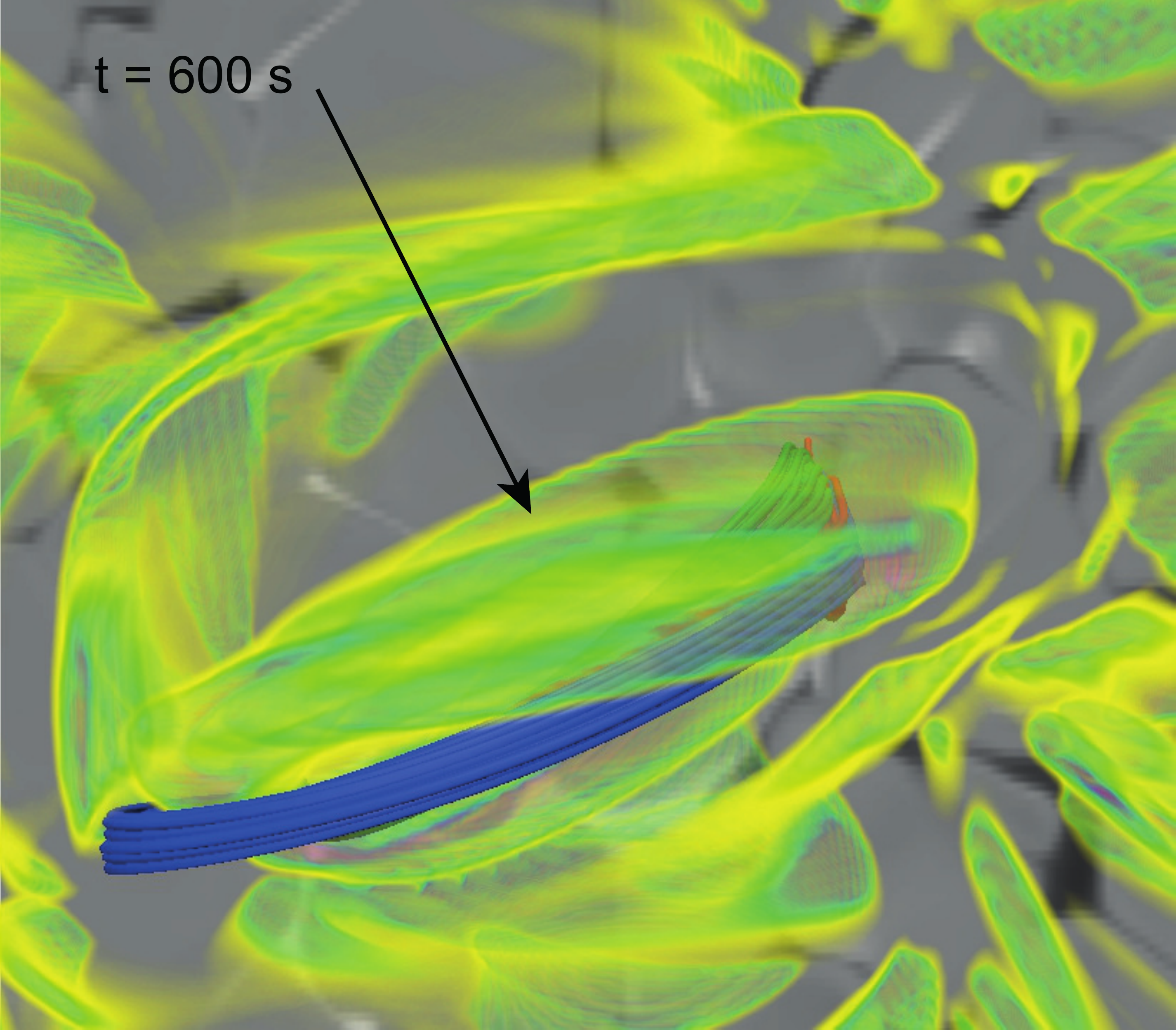}}
 \subfigure{\includegraphics[width=.30\textwidth]{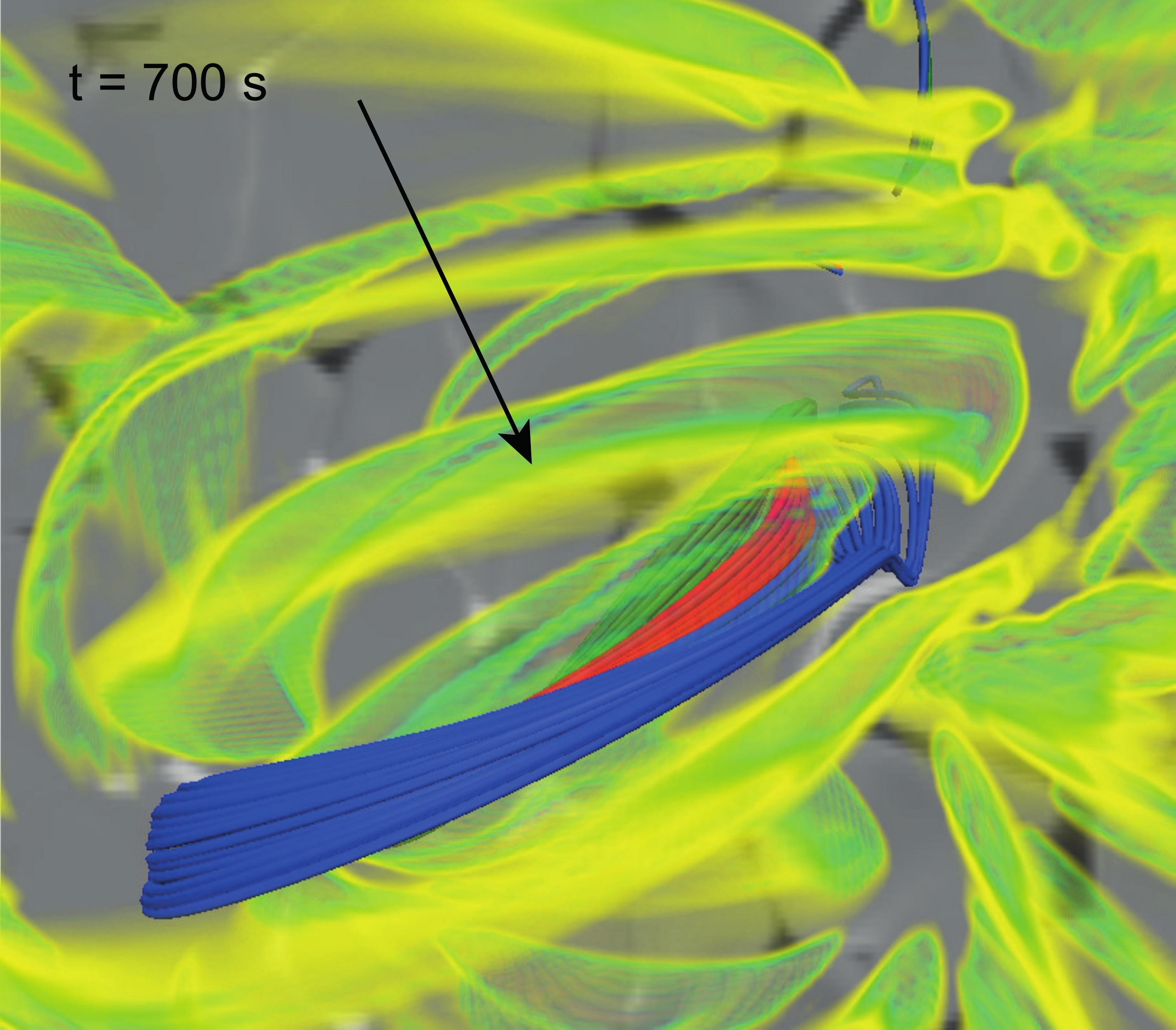}} 
   \subfigure{\includegraphics[width=.30\textwidth]{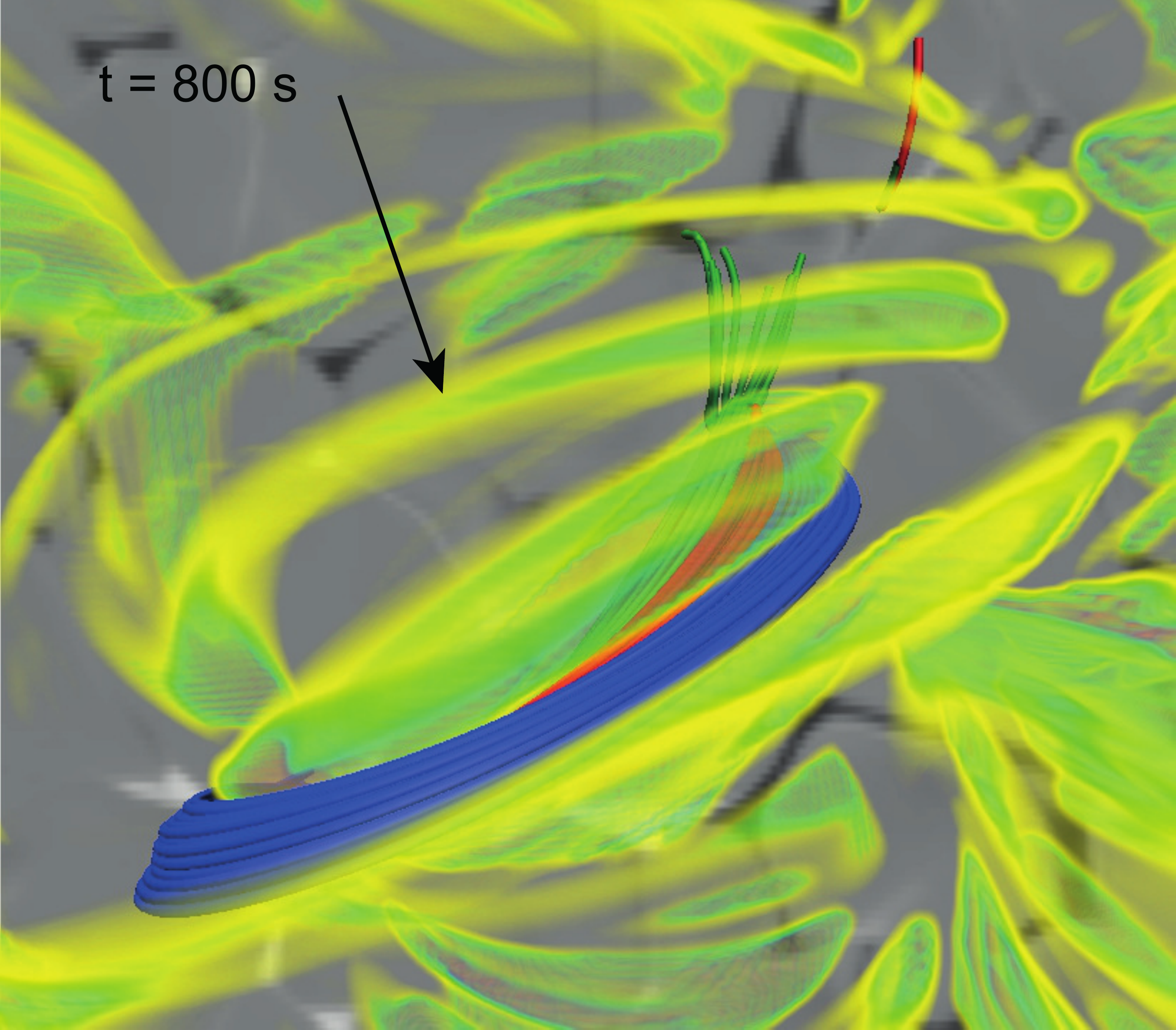}}
 \subfigure{\includegraphics[width=.30\textwidth]{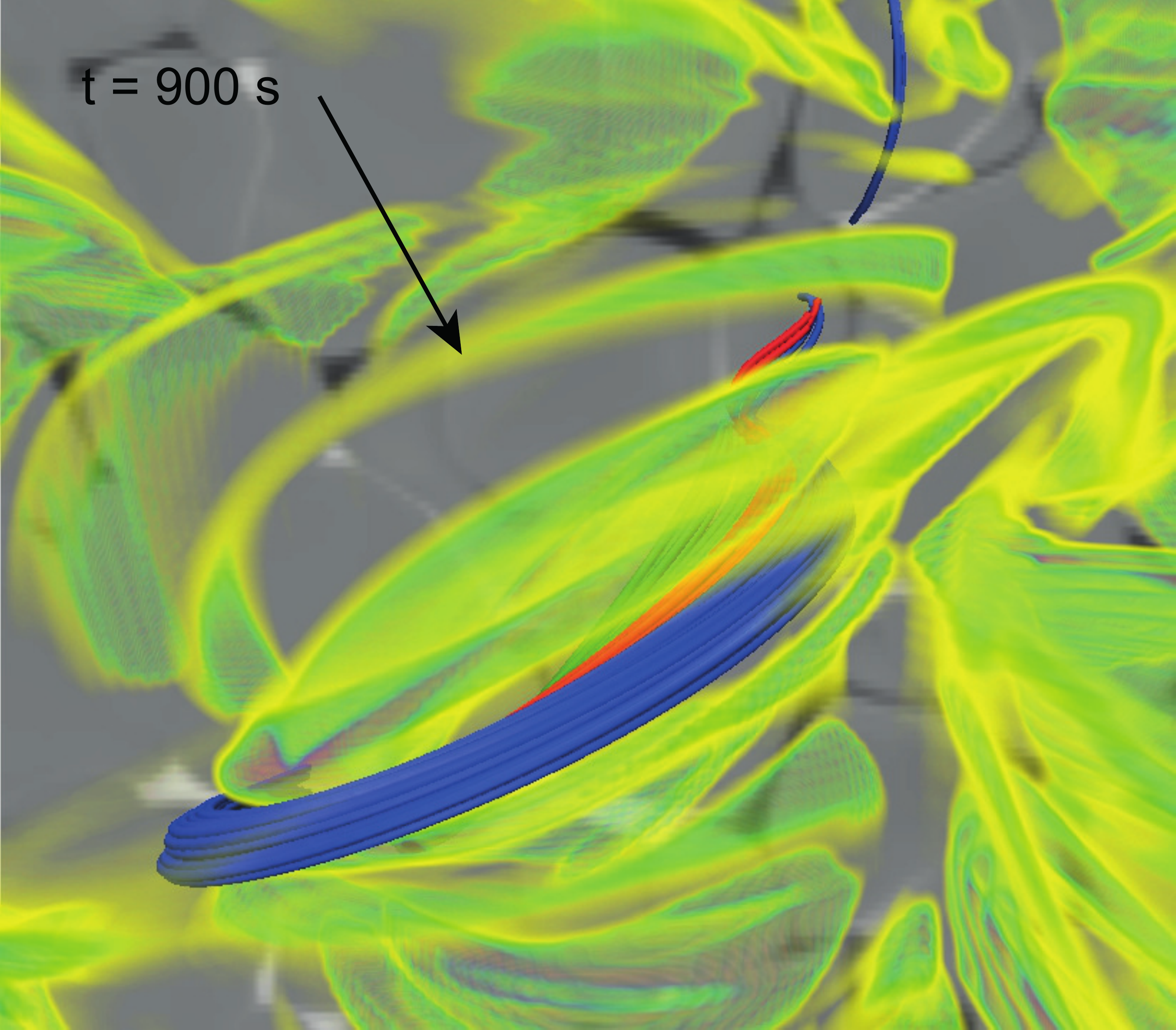}} 
 \caption{Magnetic field evolution at the Joule dissipation location shown in Figure \ref{fig:MF_angle1a} at $t=100$~s (top left), $t=200$~s, $t=300$~s (top right), $t=400$~s (middle left), $t=500$~s, $t=600$~s (middle right), $t=700$~s (bottom left), $t=800$~s and $t=900$~s (bottom right) from a top view. The panels display the evolution of the tilt in the magnetic loops and the horizontal displacement of the dissipation. The arrow points at the current sheet of interest. The panels have a $5.5\times 5.0\rm~Mm$ field of view. These representations use data from Model A.} 
 \label{fig:MF_angle3}
\end{center}
\end{figure*}

The time evolution of this (typical) current sheet is shown in figure~\ref{fig:MF_angle3} where the current density squared per particle is shown over $900$~s at intervals of $100$~s. Initially, there is little current density and the direction of field lines is nearly the same above and below the forming current sheet. After a few hundred seconds the differential angle has increased markedly as has the current density and a region of strong dissipation has formed. The dissipation region does not stay in the same place, nor is it necessarily tied to the same field lines, but moves upward in the atmosphere as well as horizontally (upwards in the series of panels shown). The upward displacement is  of the order of $1$~Mm. The horizontal displacement is most clearly visible from $t=500$~s onwards as the strongest part of the current sheet moves horizontally $\sim 1.6$~Mm with a velocity of $\sim 4~\rm km~s^{-1}$ \citep[this motion may be the equivalent of that reported observationally  by][]{1999SoPh..187..261S}. 
Note that the differential angle in the magnetic field decreases as the current sheet moves away from the region marked by field lines. Towards the end of the time series the current sheet, now located higher in the atmosphere and above the region marked by magnetic field lines in the figure, has dissipated. At the same time a new current sheet seems to be forming in the location originally stressed; the magnetic field is rooted in the photosphere and convection zone below, in which the typical time scale of plasma motions stressing the field are much longer than those found in the chromosphere and corona, allowing the repeated rebuilding of stress in a given upper atmospheric location.

\begin{figure}
\includegraphics{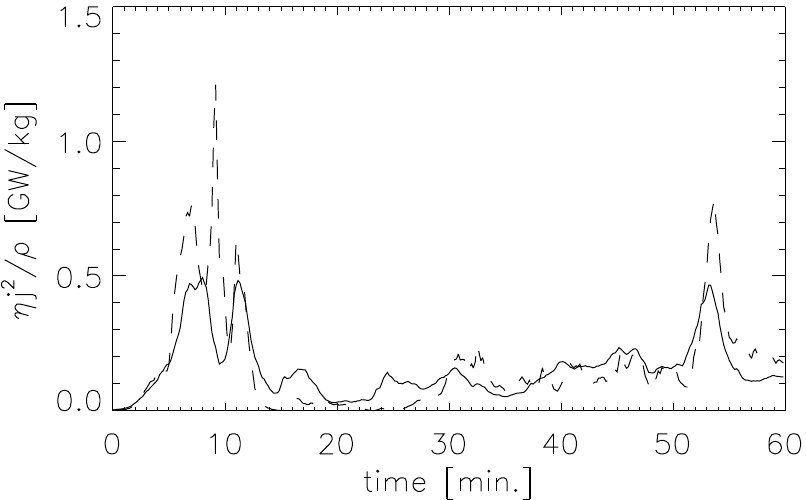}
\caption{Heating evolution in two boxes, $1\times 1\times 1$~Mm$^3$ (solid) and $\sim 0.2\times \sim 0.2\times \sim 0.2$~Mm$^3$ (dashed)  at the top of the loop shown in Figure~\ref{fig:MF_angle1a}. Data from Model A.} 
\label{fig:twist_loc}
\end{figure}

At certain locations, such as the one described above, Model A shows recurrent behavior, or periodicities, in the dissipation. This is shown in figure~\ref{fig:twist_loc}. We select a fixed volume of $1\times 1\times 1$~Mm$^3$ around the top of the current sheet studied and integrate the total heating in this volume for the entire time series. Dissipation is initially at a very low level as photospheric motions have not yet had time to build up sufficient stresses in the chromospheric and coronal field. After some 10~minutes we find a period of strong (and repeated) dissipation; this is the same series of current sheet formation and dissipation that was described above in figures~\ref{fig:MF_angle1a} and \ref{fig:MF_angle3}. After this series of large events we find smaller maxima after 30 and 45~minutes, before a  very large dissipation event after 53~minutes. Each individual dissipation event lasts some few hundred seconds. Both complex and simpler temporal structures of the maxima are found,
  with either a single maximum or  repeated maxima separated in time by some hundred seconds.

In Model A the magnetic topology is such that we find that a number of field lines are open and pierce the upper boundary. In addition, there are also several field lines that are ``quasi-open''  in the sense that 
they pass through the upper corona several times as a result of the horizontal periodic boundary condition forming very long loops before returning to the lower atmosphere and photosphere.

\begin{figure}[htbp]
\begin{center}
 \subfigure{\includegraphics[width=.49\textwidth]{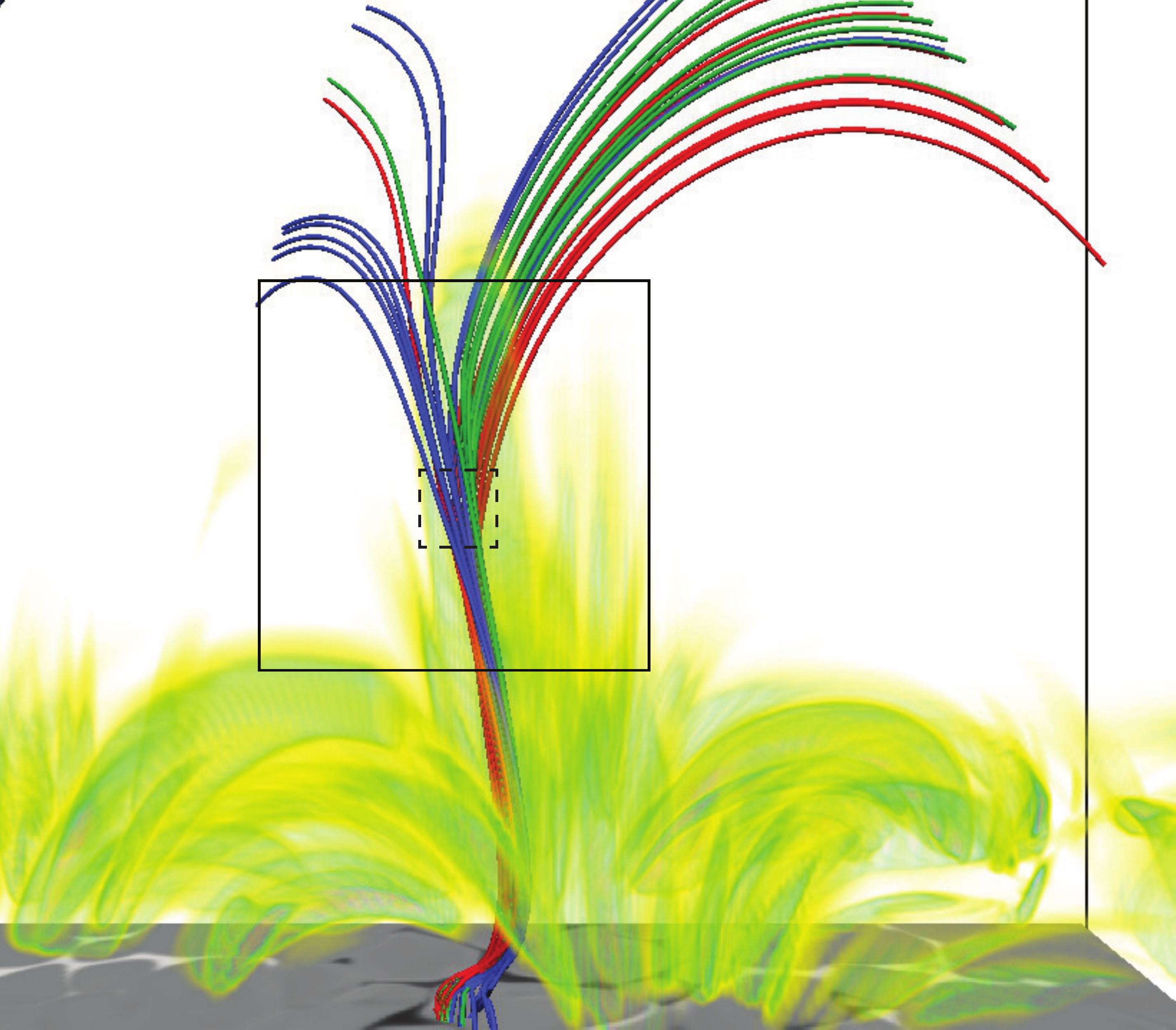}}
  \subfigure{\includegraphics[width=.49\textwidth]{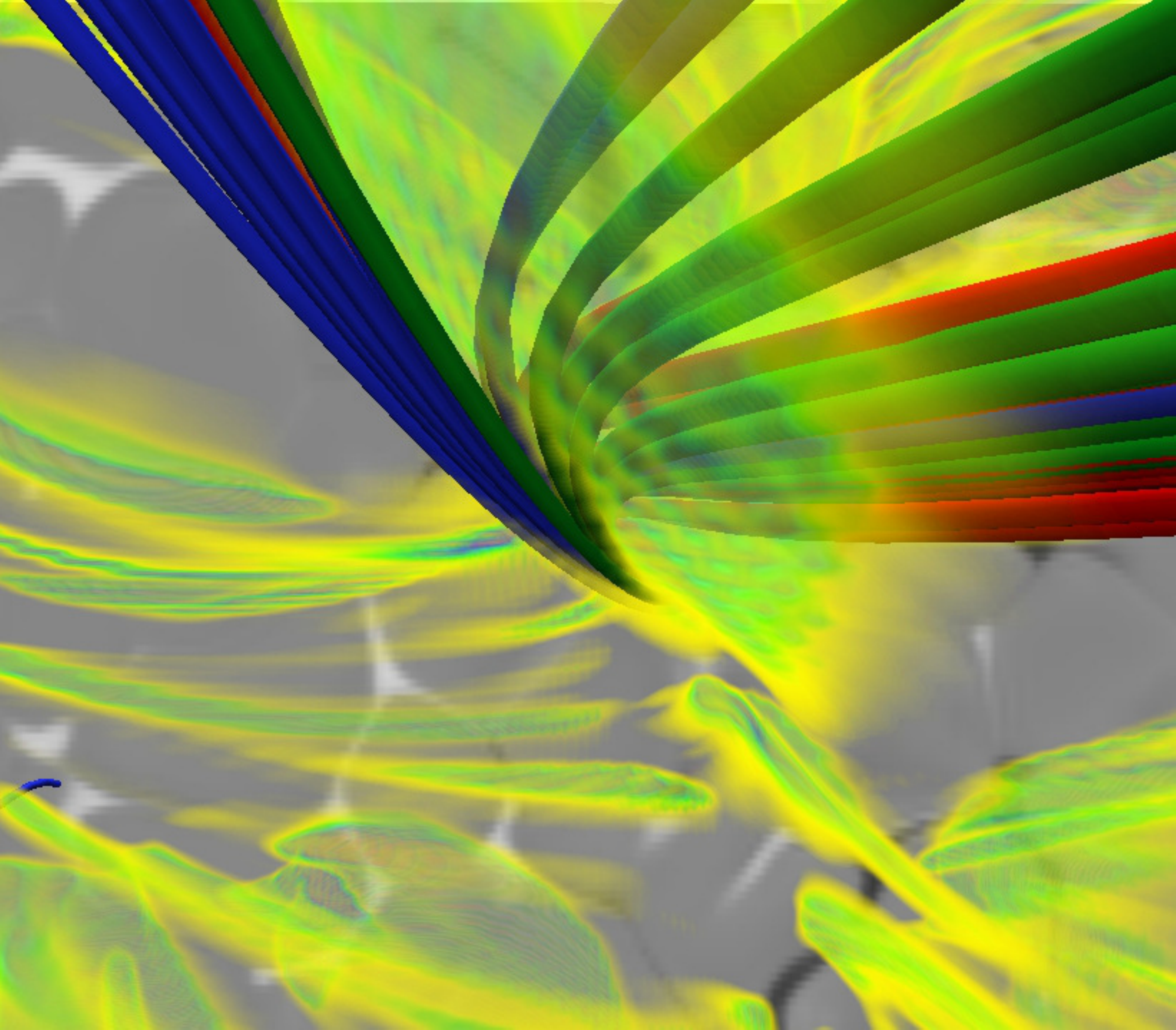}}
 \caption{Magnetic field topology surrounding a dissipative ``finger'' extending well into the corona along a long ``open'' loop in the low resolution model at $t=500$~s. The current density squared per particle ($\left( j^{2}/\rho\right)$) is shown using the same color table as in Figure \ref{fig:JH_Model A_Model B}. Also shown are the approximate locations of the two fixed boxes, size $1\times 1\times 1$~Mm solid line, size $0.2\times 0.2\times 0.2$~Mm dashed line) used to follow the heating evolution in area in Figure~\ref{fig:twist_open_loc}. Note the twist lower down, and the fan-like structure towards the top. The right panel field of view is $3.0\times 2.7~\rm Mm$. Data from Model A.} 
 \label{fig:MF_angle2}
\end{center}
\end{figure}

As mentioned above we also find significant heating at these locations that could be similar to the fan shaped { quasi-separatrix layers (QSL)} regions \citep[see][]{1995JGR...10023443P} seen in TRACE observations as discussed by \citet{2010ApJ...719.1083S}. In figure~\ref{fig:MF_angle2} the current density per particle is shown for such a region, at time $t=500$~s, showing a ``finger'' of large heating extending to some 7~Mm above the photosphere, much higher than for the closed loops found in regions where the field is more horizontally oriented. When visualizing the magnetic field lines in the vicinity of the heating event we see that the field appears twisted, and in addition, that at the greatest heights the field spreads out in a fan shaped structure. Similar to what we found for the closed loop, the heated region moves of order 1~Mm during a 1000~s period and the heating remains high as long as the field lines are twisted; when the twist dissipates the heating is reduced regardless of whether or not we see a fan shaped structure in the field. 

\begin{figure}
\includegraphics{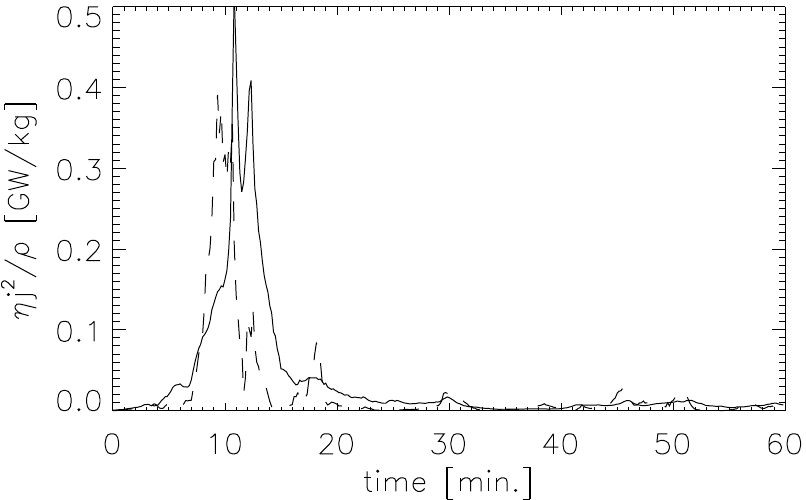}
\caption{Heating evolution in a $1\times 1\times 1$~Mm$^3$ volume (solid) and $\sim 0.2\times \sim 0.2\times \sim 0.2$~Mm$^3$ (dashed) at the center of the twist as shown in Figure~\ref{fig:MF_angle2}. Data from Model A.} 
\label{fig:twist_open_loc}
\end{figure}

The heating evolution of this region as a function of time in a $1\times 1\times 1$~Mm$^3$ volume centered on the twist seen in figure~\ref{fig:MF_angle2} is shown in figure~\ref{fig:twist_open_loc}.

Heating events in Model B show the same general tendency as found in Model A: heating events in closed loops are associated with small angle changes in the magnetic field direction that last of order 100~s before they dissipate as the field becomes more nearly parallel. The location of greatest heating moves with the general plasma flow as field lines are advected by the photospheric flow. In most, if not all cases, we do not see any large changes in the topology of the field or obvious bi-directional jets associated with the reconnection that dissipates the stresses built up by photospheric motions. The field in long, or open, loops that are strongly heated shows evidence of twist and often spreads out in fan shaped structures.

\subsection{LOCAL DISSIPATION EVENTS}

\begin{figure*}
\includegraphics{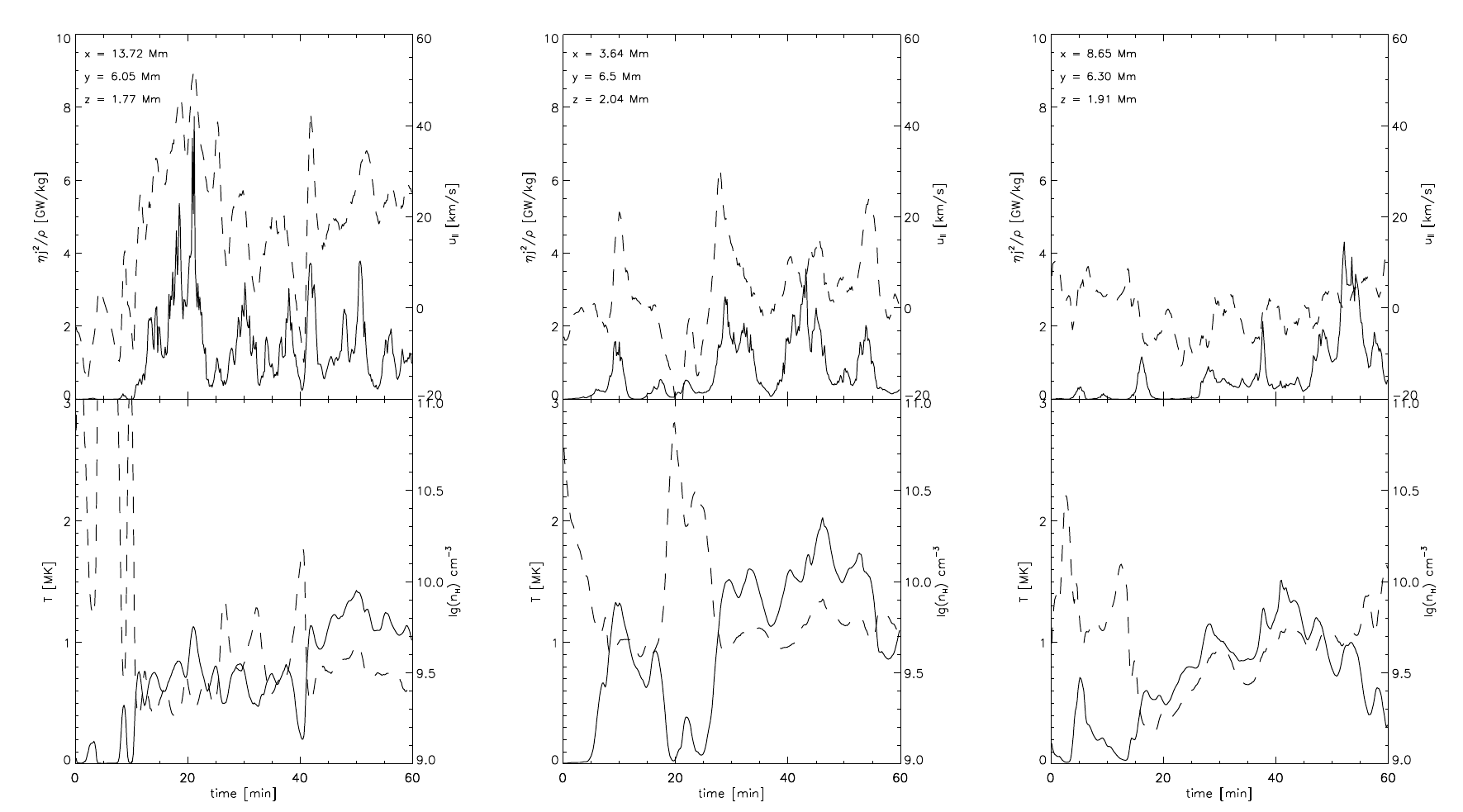}
 \caption{[Top] Joule heating per unit mass (solid line) and  velocity along the magnetic field (long dashed line), [bottom] temperature (solid line) and particle density (long dashed line) as a function of time for the entire run of Model A at three locations with high dissipation in the low corona/transition region.} 
 \label{fig:lowcor_heat_inten}
\end{figure*}

\begin{figure*}
\includegraphics{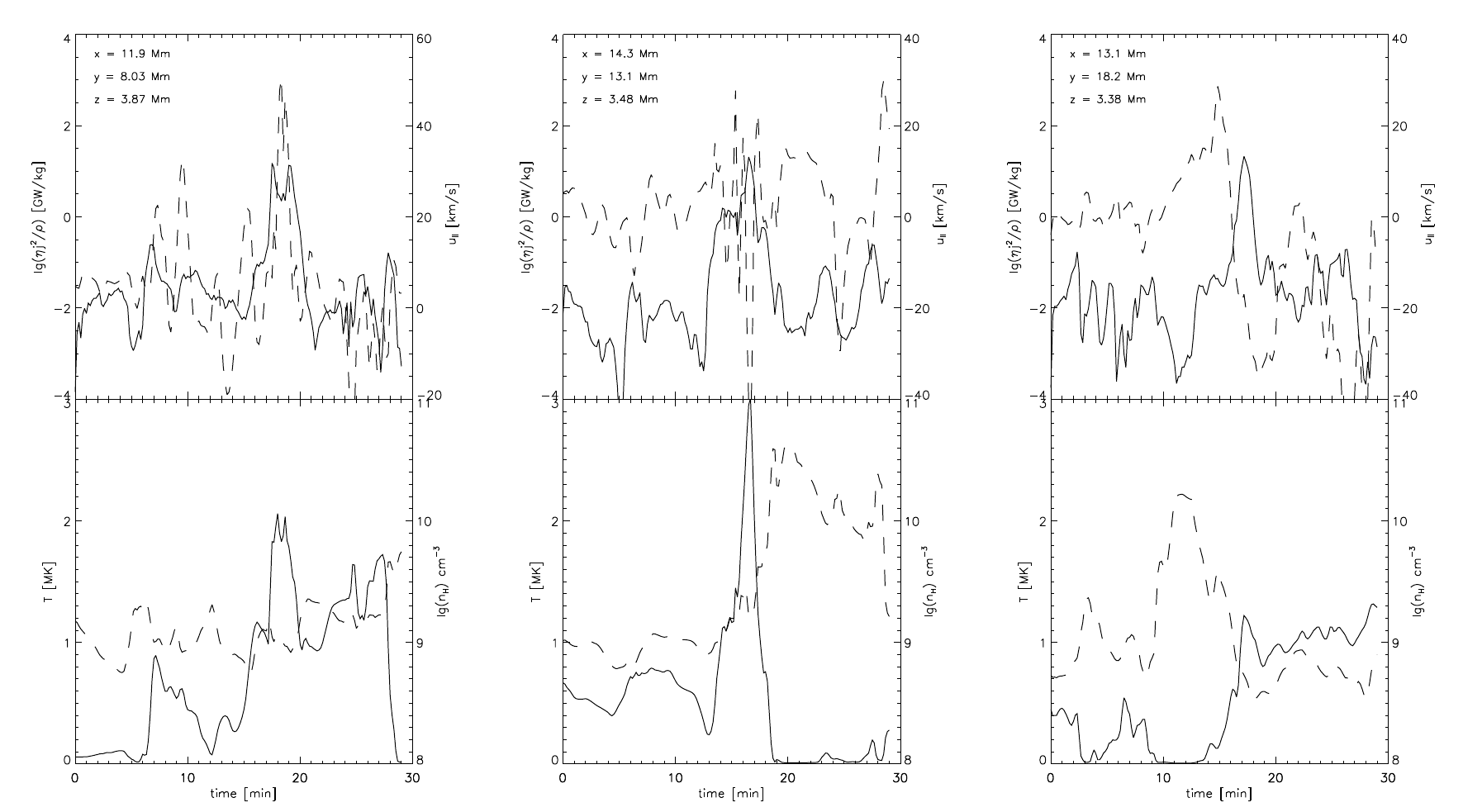}
 \caption{Same as Figure \ref{fig:lowcor_heat_inten}, but for Model B.} 
 \label{fig:lowcor_heat_inten_hr}
\end{figure*}

Let us now consider heating events at specific locations in the region of most effective heating per particle, i.e., in the vicinity of upper chromosphere, transition region, and lower corona. As already alluded to in figures~\ref{fig:twist_loc} and \ref{fig:twist_open_loc} the heating in a given location will be episodic, but sometimes recurring. We have searched for sites of greatest heating per particle, and in figures~\ref{fig:lowcor_heat_inten} and \ref{fig:lowcor_heat_inten_hr} we show three distinct locations where vigorous dissipation occurs in Model A and B, respectively. We selected these locations by averaging the Joule dissipation per particle over the entire time series and choosing the locations (cells) with the highest average, excluding locations neighboring other strong heating cells.

The dissipation per particle in fixed volumes of roughly $195\times 195 \times \sim 200$~km centered on the grid cell 
with the highest average value are then found. The same methodology was used for both Model A and B. Even though the volumes integrated are roughly the same in both models the results are significantly different as discussed below.

In model A, the three strongest events are all found at heights between $z=1.5$~Mm and $3$~Mm, on average at $2$~Mm above the photosphere. While all locations show large variations in the heating rate, all have significant heating during most of the model run. As in the examples of the closed loop discussed in the previous section the heating rate shows peaks lasting some $100-200$~s that recur on average $9$ times during the hour, or every $6-7$ minutes, though with varying peak intensity. The lifetime as measured in this way is clearly a combination of the co-moving lifetime of the event and the time it takes the dissipation region to cross the measurement volume. For reference, the average plasma velocity in this region of the atmosphere, $2$~Mm above the photosphere, 
 is of order $10\rm~km~s^{-1}$ so a $200$~km wide structure should take roughly $30$~s to pass through a volume with the dimensions chosen here.

Figures~\ref{fig:lowcor_heat_inten} and \ref{fig:lowcor_heat_inten_hr} also display the velocity along the magnetic field, and in the lower panels, the temperature and the density. The velocity along the magnetic field shows peaks that seem to be correlated with the heating events, {presumably caused by the rapid expansion caused by a drastic heat input, but possibly also due the sudden pairing of two flux tubes with significantly different thermodynamic properties.}
The temperature and density also seem to be correlated with the vigorous heating events, however the correlation is not nearly as clear as for the velocity, and while the temperature generally increases with increased heating, the density can either increase or decrease in response to an increased heating rate.

While large heating per particle in Model A is concentrated near $2$~Mm, in Model B the average height of the largest events is found to be nearly $3.5$~Mm, spanning a range from $2.9$~Mm to $3.9$~Mm. Heating events in Model B also appear much more sporadic than in Model A and the peaks are narrower in time, as can be seen in figure \ref{fig:lowcor_heat_inten_hr}, where the heating is displayed in a logarithmic scale. On the other hand the peak values are comparable, $\simeq 50\rm~MW~kg^{-1}$ in both cases. As noted earlier, the heating in Model B is quite intermittent and therefore there is no evident recurrence of heating at the locations of the volumes chosen. This large intermittency could be due to increased fragmentation of current sheets allowed by the better spatial resolution of Model B \citep[see][]{K.Galsgaard061996}.

The velocity, temperature and density behavior is similar to that shown in Model A. The maximum absolute values of the velocity along the line seems to show a good correlation with heating events in these locations. 

A somewhat better estimate of heating event lifetimes in a given location can be found by calculating the auto-covariance of the dissipation time series at that location. We can then use the derived auto-covariance against the lag to determine the duration of the dissipative events in each cell. Here we define the lifetime of a dissipative event at a given location to be equal to the width of the auto-covariance curve at half maximum. We considered a lag of up to $400$~s which is longer than the estimated lifetime of the large scale dissipative structures. This choice of lags proved to be reasonable since the number of localized heating events that last longer than $400$~s is negligible in our models.  

\begin{figure*}
\includegraphics{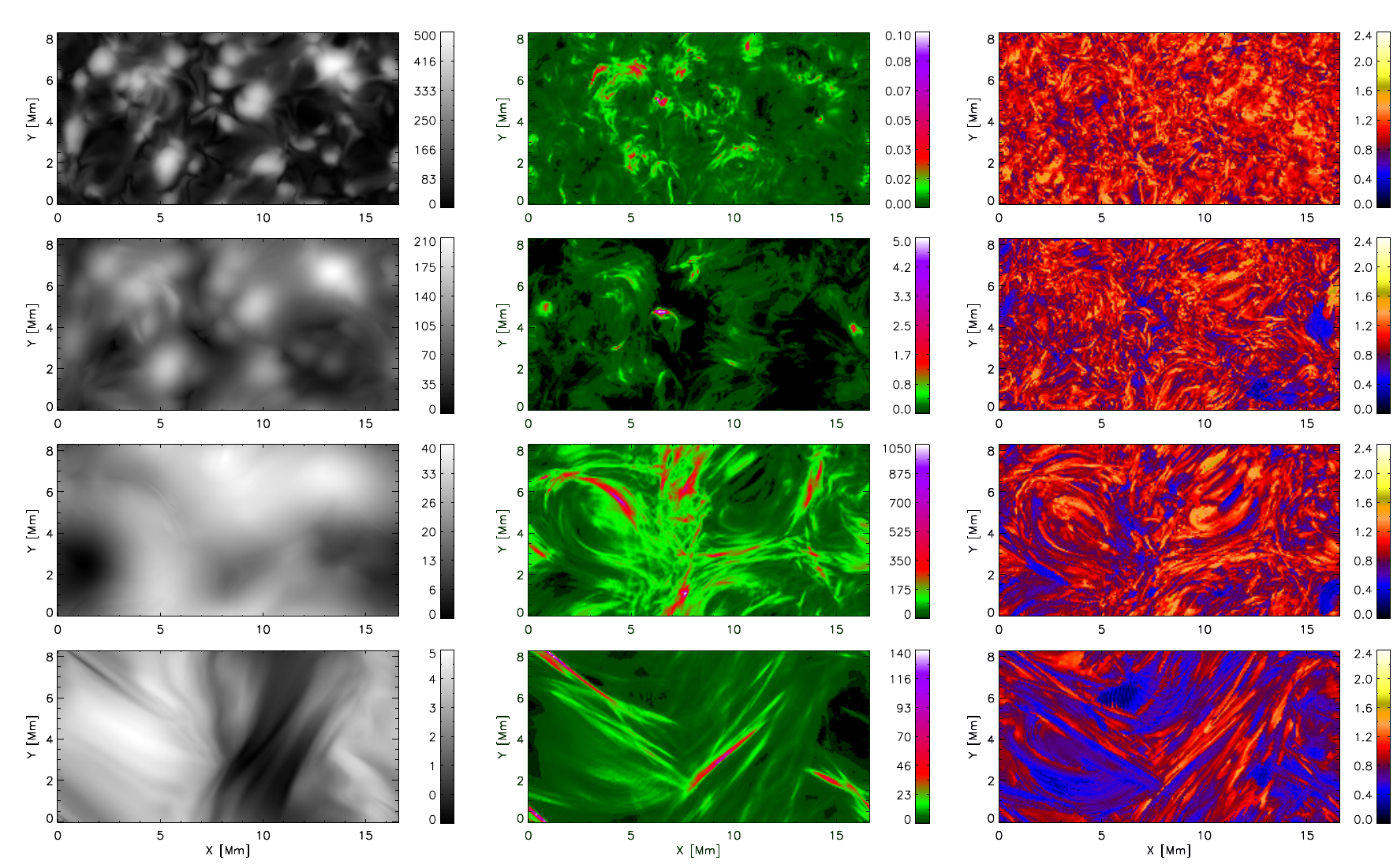}
 \caption{Magnetic field [G] at $t=2000\rm~s$ (left), average Joule dissipation over the time series $\left[\rm kW~kg^{-1}\right]$ (middle) and  logarithm of the auto-covariance time scale [s] (right) at $z=0.490$, $z=1.002$, $z=2.992$ and $z=9.244$ Mm (from the top to the bottom) for Model A.} 
 \label{fig:time_delay_maps}
\end{figure*}

\begin{figure*}
\includegraphics{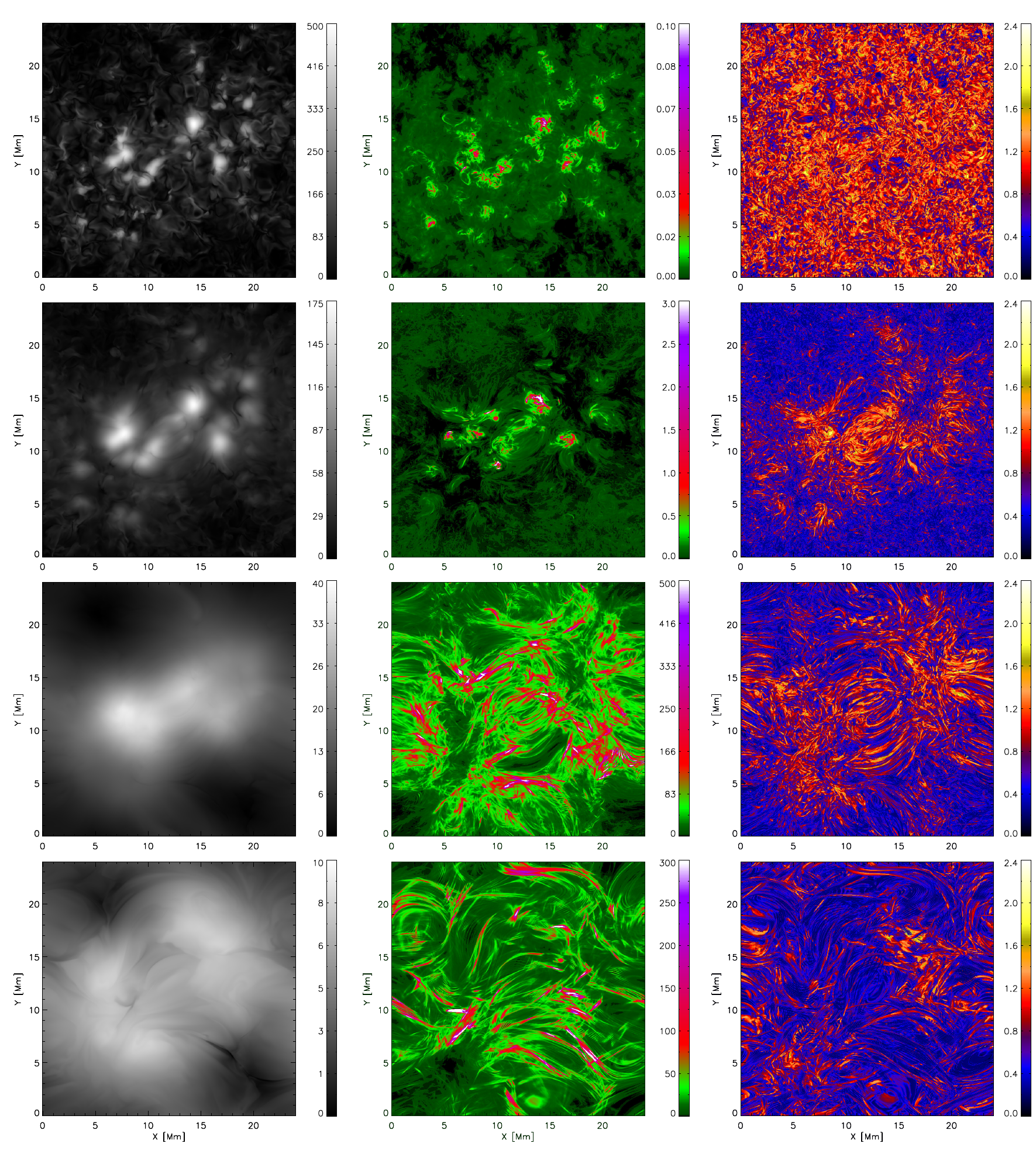}
 \caption{Same as figure~\ref{fig:time_delay_maps}, but for Model B.} 
\label{fig:time_delay_mapshr}
\end{figure*}

Figures~\ref{fig:time_delay_maps} and \ref{fig:time_delay_mapshr} show maps of the auto-covariance time scales at four different heights in the atmosphere for Model A and B, respectively: in the low chromosphere $\left(490\rm ~km\right)$, in the middle chromosphere $\left(1\rm ~Mm\right)$, in the transition region/lower corona $\left(3\rm ~Mm\right)$, and in the corona $\left(9.2\rm ~Mm\right)$. These time scales are computed using a cell size of $65\times 65\times \sim 40 ~\rm km^{3}$ for Model A and $31\times 31\times \sim 50 ~\rm km^{3}$ for Model B. Also shown are 
the average heating rates per particle and the unsigned magnetic field strength at $t=2000~\rm s$ for the same heights.

The lower chromospheric auto-covariance time scale map is very highly structured 
and does not bear any clear relation to the map of average Joule heating per unit mass in the same region. The time scale for heating events at this height is of order a minute or longer (see below). Regions of long-lived heating form roughly elliptical or circular structures with dimensions of order 1~Mm or smaller. The map of the time scales in the middle chromosphere is very similar to that found 500~km below, though at this height the elliptical long duration structures are perhaps somewhat more eccentric. Again, there is no clear correlation between the time scale map and the average heating per unit mass map. Perhaps this is due to the dominance of acoustic shocks at these high-to-medium plasma-$\beta$ heights, where wave motions push and deform fields, increasing the Joule heating rate as oppositely directed fields are brought together at the whim of plasma dynamics.

In the low plasma-$\beta$ transition region, 3~Mm above the photosphere, we see a marked change in the structure of the auto-covarience time scale map: the general impression is that structures are more ordered by the magnetic field and that long duration regions are larger. These regions are clearly more oblong at this height and now bear some relation to the structure defined by the average heating rate. In the corona, at 9.2~Mm,  long duration structures mainly form linear shapes with lengths of up to at least 5~Mm, though there is also a large nearly circular structure visible.  The relation to the average heating map which shows similar linear structures in the same locations is quite evident.

The time scale maps of Model B, figure~\ref{fig:time_delay_mapshr}, are quite similar to those found for Model A, but in general we find that Model B is characterized by shorter time scales and, of course, more highly structured shapes. This could be due to the higher resolution in the Model B simulation that allows the heating events to fragment into intermittent smaller scale structures. At this resolution the correlation between the average heating map and the time scale map seems better, extending even down to the middle chromosphere. In the transition region and corona the relation is quite obvious: regions of large heating are also regions of longer lifetimes. Again, we find circular or slightly elliptical shapes in the chromosphere, and much more linear structures in the transition region and corona.

As discussed above it should be clear that the timescale derived will be a combination of the timescale for a given structure to cross our chosen volume and the intrinsic co-moving lifetime of the heating event itself (very short timescales can occur if there is very little variation in the heating rate --- in that case the auto-covariance timescale is determined by the small random fluctuations), possibly modified by the occurrence of two or more heating regions in the same volume. We can shed some light on this issue by varying the size of the grid cells utilized and repeating the analysis described above. Since we are computing the auto-covariance for each cell in the simulated atmosphere there are cells where the variations of the heating are very small; i.e. ``background cells''. This implies a very slow decrease of the auto covariance against the lag curve. Since the width at half height of this curve is the parameter that gives the lifetime of the events, these background cells will contribute with an overestimate of the heating events lifetime. The average values found are therefore an upper limit for event lifetimes. 

\begin{figure*}[t]
\begin{center} 
 \subfigure{\includegraphics[width=1\textwidth]{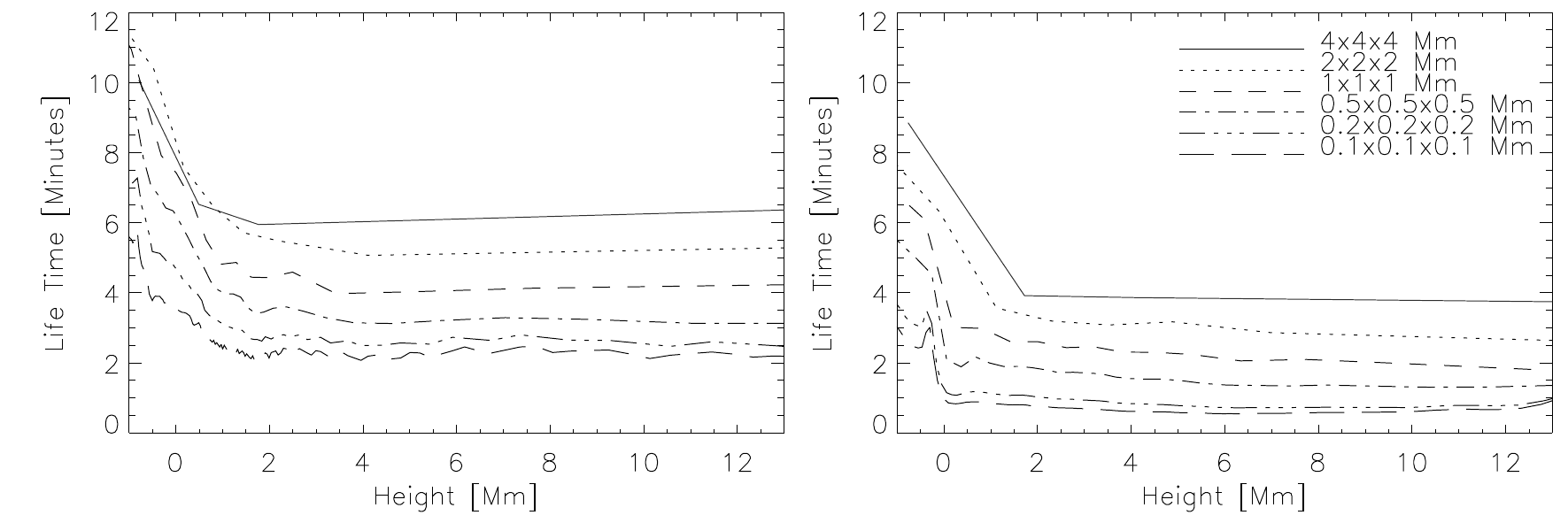}}
 \caption{ Average lifetime of the heating events, as measured by the auto-covariance (see text), against height for Model A (left) and Model B (right). Increasing  the size of the cell considered in the analysis increases the measured average lifetime.} 
 \label{fig:ave_ltimes_lr}
\end{center}
\end{figure*}

We have rebinned the computational box with cells of different dimensions, keeping the same physical dimension for both models.  Figure \ref{fig:ave_ltimes_lr} displays the average ``width at half height'' of the co-variance function, {\it i.e} the dominant lifetime, versus height for the two models, for cells of different physical dimensions. For both models the lifetime increases with physical dimension, presumably as the volume considered grows to encompass the region containing the entire lifetime of any heating event and/or several heating events. It is clear that in Model A the average time versus height is greater for all cell dimensions and at all heights, though the difference is smaller in the vicinity of the photosphere.

Based on figure~\ref{fig:ave_ltimes_lr} we find that dissipative events have lifetimes between $2$ and $7$~minutes in Model A and between $1$ and $5$~minutes in Model B. If there is only one event per considered box size then these lifetimes should give upper limits to the total co-moving lifetime of a single event. This estimate assumes that, using a typical speed of $<10\rm~km~s^{-1}$, dissipative events cannot move further than $4$~Mm ($4200$~km) in $7$~minutes. Thus, the time scale found when considering cells of $4\times 4\times 4$~Mm$^3$ should be the greatest possible timescale for a single event. This gives upper limits of co-moving dissipation lifetimes of $7$~minutes for Model A and 4 minutes for Model B, in the transition region and corona. Examining even larger boxes would imply that one is looking at several events and the timescale will increase accordingly, asymptotically tending towards infinity as the number of events grows, this is most probably the case for the largest boxes of $4\times 4\times 4$~Mm$^3$.

\section{DISCUSSION AND CONCLUSION}

We present two models with fully self consistent coronal heating, and analyze the characteristics of this heating in the transition region and corona. The velocity pattern in the photosphere --- granulation and associated flows --- is driven by self-consistent convection, modeled to several megameters below the photosphere, rather than being given by a statistical description as done by \cite{B.Gudiksen012005,S.Bingert062011}. In addition the models discussed here have significant improvements in spatial resolution compared with these previous models. 
As time passes, magnetic fields are braided by photospheric and convective motions and gradients build up in the chromospheric and coronal fields in the form of current sheets. We find significant Joule heating in the vicinity of the current sheets, and this heating comes to dominate the energetics in the low plasma-$\beta$ regions of the atmosphere, i.e. at heights greater than some 1~Mm above the photosphere up into the high corona.

The general pattern of heating found in the two models presented here is very similar, but Model B has significantly greater heating in the photosphere and somewhat greater in the upper atmosphere, even though the average unsigned magnetic field is roughly the same.
In both models the heating we find is roughly proportional to the magnetic field energy density, but the topology of the field likely also plays an as yet not wholly determined role. The distribution of the magnetic field in the photosphere determines the magnetic field scale height, with closely spaced magnetic polarities of opposite sign giving small scale heights and widely separated polarities of opposite sign giving larger scale heights. This in turn determines the location of the maximum heating per particle, given by the ratio of the magnetic and plasma scale heights in the chromosphere. The fall off of heating per particle with height in the corona will also be smaller the greater the magnetic scale height; based on this argument we expect unipolar plage regions to still peak in the upper chromosphere and transition region, but also have high heating rates extending to large heights. 
The results presented here are therefore probably most representative of the coronae above the quiet sun or enhanced network. For these ``normal field configuration'' regions, the simulations show that the Joule heating events are most effective in the vicinity of the upper chromosphere, transition region, and lower corona.

On average we find that the corona takes a long time to heat and fill with mass, and the question could be raised as to whether a coronal loop ever reaches an equilibrium, i.e. that the coronal mass is consistent with the average heating rate, given such a strongly time-and-space varying, impulsive heating \citep[see e.g.][]{S.Bradshaw022010,2011ApJ...734...90W}. The total energy input into the corona is fairly constant over one hour time scales simulated here. On the other hand, the heating is spatially intermittent and temporally episodic. Likewise, the Poynting flux injected into the corona varies strongly with time, but on average gives a consistent, and as compared to estimates based on observations, reasonable energy flux into the corona. The highly variable nature of the Poynting flux and the necessity of measuring all the vector components of the velocity and magnetic field indicates that it may be difficult to measure the heat flux injected into the upper atmosphere based on direct photospheric observations \citep{2015PASJ..tmp..156W}. 

Spatially, the heating is concentrated in current sheets, the thicknesses of which are set by the numerical resolution, that stretch along loops. Some current sheets fragment with time, spawning smaller current sheets as they are driven to smaller scales by photospheric forcing motions: i.e. small heating events are aligned in long strands, oriented along the magnetic field lines and associated with small changes in the angle of the magnetic field over the current sheet where heating is concentrated. These strands therefore show loop like structure with lifetimes of the order of $100$~s. This tendency towards fragmentation and intermittent heating seems more pronounced in the higher spatial resolution Model B simulation.

Magnetic field maps are not a good guide to locations of large heating; there is much more structure in maps showing regions of high heating or maps of heating lifetimes than in maps of the magnetic field. 
Localized heating events have lifetimes that are about $2$~minutes for Model A and about $1$~minute for Model B. The heating events seem to show some recurrence. The intrinsic co-moving lifetime of  events is estimated to be smaller than $5$~minutes.

In a given location, as a function of time, heating is highly episodic and the velocity, density, and temperature 
of the plasma will respond on the same timescales as well as on the slower timescales dictated by the processes cooling the plasma: e.g. enthalpy, flows, conduction, and radiative losses. The next step is to construct synthesized observations that clearly link heating events to the underlying processes occurring in the chromosphere and corona. This may help to clear up the the contradictory observational results shown by several authors  
 \citep{S.Krucker071998,D.Berghmans081998,C.Parnell012000,M.Aschwanden062000,A.Benz2002}. A follow up study to pursue the connection between episodic coronal heating and its observational signatures in upper chromospheric, transition region, and coronal diagnostics is under way.

\acknowledgments{This research was supported by the European Commission funded Research
Training Network SOLAIRE. It was also supported by  the Research
Council of Norway through grants of computing time from the Norwegian
Programme for Supercomputing and by the Swiss National Science 
Foundation under grant n$^o$ 153302.  The research leading to these results has received funding from the European Research Council under the European Union's Seventh Framework Programme (FP7/2007-2013) /
ERC Grant agreement n$^o$ 291058 and support by NASA grants NNX08AH45G, NNX08BA99G, NNX11AN98G, NNM07AA01C (Hinode), and NNG09FA40C (IRIS).
To analyze the data we have used IDL and Vapor (http://www.vapor.ucar.edu).}


\end{document}